\documentclass[aps,prx,reprint,twocolumn,notitlepage, superscriptaddress]{revtex4-1}
\usepackage{graphicx}
\usepackage{amsmath,amssymb}
\usepackage{color}
\usepackage{bm}
\usepackage{mathtools}
\usepackage{hyperref}
\usepackage{soul}

\newcommand{\secref}[1]{Sec.\,\ref{#1}}

\begin{document}

\title{Spontaneous breaking of U(1) symmetry at zero temperature in one dimension}
\author{Haruki Watanabe}\email{hwatanabe@g.ecc.u-tokyo.ac.jp}
\affiliation{Department of Applied Physics, The University of Tokyo, Tokyo 113-8656, Japan}

\author{Hosho Katsura}\email{katsura@phys.s.u-tokyo.ac.jp}
\affiliation{Department of Physics, 
The University of Tokyo, Hongo, Bunkyo-ku, Tokyo 113-0033, Japan}
\affiliation{Institute for Physics of Intelligence, The University of Tokyo, Hongo, Bunkyo-ku, Tokyo 113-0033, Japan}
\affiliation{Trans-scale Quantum Science Institute, The University of Tokyo, Hongo, Bunkyo-ku, Tokyo 113-0033, Japan}

\author{Jong Yeon Lee}\email{jongyeon@illinois.edu}
\affiliation{Department of Physics, University of California, Berkeley, California 94720, USA}
\affiliation{Department of Physics, University of Illinois at Urbana-Champaign, Urbana, Illinois 61801, USA}

\date{\today}

\begin{abstract}
The Hohenberg--Mermin--Wagner theorem states that there is no spontaneous breaking of continuous symmetries in spatial dimensions $d\leq2$ at finite temperature. At zero temperature, the classical/quantum mapping further implies the absence of continuous internal symmetry breaking in one dimension, which is also known as Coleman's theorem in the context of relativistic quantum field theories. 
One route to violate this ``folklore'' is requiring an order parameter to commute with a Hamiltonian, as in the classic example of the Heisenberg ferromagnet and its variations. 
However, a systematic way of understanding the spontaneous breaking of internal U(1) symmetries has been lacking.
In this Letter, we propose a family of one-dimensional models that display spontaneous breaking of a U(1) symmetry at zero temperature,  although the order parameter does not commute with the Hamiltonian, unlike the Heisenberg ferromagnet. We argue that a more general condition for this behavior is that the Hamiltonian is frustration-free.
\end{abstract}

\maketitle

\emph{Introduction.---}
In the study of condensed matter physics, understanding symmetry and how it can be spontaneously broken stands as a foundational pillar underpinning the elucidation of a diverse spectrum of emergent phenomena.
A pivotal theorem in this domain, known as the Hohenberg--Mermin--Wagner (HMW) theorem~\cite{Hohenberg,PhysRevLett.17.1133}, states that any continuous symmetry cannot be spontaneously broken in spatial dimension $d\,{\leq}\,2$ at a finite temperature $T\,{>}\,0$. The theorem has been recently extended to higher-form~\cite{EthanLake} and multipole symmetries~\cite{PhysRevB.105.155107,Kapustin},  offering insights into highly constrained systems. With quantum-classical mapping~\cite{Sondhi1997}, the theorem further implies the absence of continuous symmetry breaking in one-dimensional quantum systems at zero temperature.

Notably, an exception has been known to the $T\,{=}\,0$ version of the HMW theorem: the $\textrm{SO}(3)$ symmetric Heisenberg ferromagnet and its variations~\cite{anderson2018basic,PhysRevLett.99.240404,10.21468/SciPostPhysLectNotes.11,WatanabeARCMP}. At a conceptual level, this exception has been understood to arise from the fact that an order parameter commutes with the Hamiltonian as it is simultaneously the generator of the global symmetry, which is possible only for non-abelian symmetry.  
However, it was recently shown that a family of frustration-free Hamiltonians naturally occurring in the description of dynamical quantum systems may exhibit spontaneous breaking of the abelian $\textrm{U}(1)$ symmetry in all dimensions~\cite{JongYeonLee}.
%

In this Letter, we propose and examine a new series of one-dimensional spin models that display spontaneous breaking of U(1) symmetry at $T\,{=}\,0$, whose order parameter \emph{does not} commutes with the Hamiltonian. Then, we argue that the general criterion for this behavior is that the Hamiltonians is \emph{frustration-free}: The Hamiltonian $\hat{H}=\sum_{i=1}^L\hat{H}_i$ is frustration-free if and only if there exists a simultaneous eigenstate of $\hat{H}_i$ with its lowest eigenvalue $\alpha_i$ for every $i$~\cite{TasakiBook}\footnote{Without loss of generality, we set $\alpha_i=0$ by properly choosing the constant in $\hat{H}_i$.}. In other words, the ground state of $\hat{H}$ simultaneously minimizes all $\hat{H}_i$'s, although $\hat{H}_i$'s do not have to commute with each other. 
Leveraging the understanding that excitations in gapless frustration-free systems are softer than linearly dispersive modes~\cite{GossetMozgunov,Anshu}, we demonstrate that frustration-free systems can bypass the constraints of the HMW theorem. 

\emph{Ferromagnetic Heisenberg Model.---}
Let us begin by briefly reviewing the ferromagnetic Heisenberg model for spin-$s$ spins. The Hamiltonian is $\hat{H}^{\text{(FM)}}\coloneqq\sum_{i=1}^L\hat{H}_i^{\text{(FM)}}$ with
\begin{align} \label{eq:FM}
\hat{H}_i^{\text{(FM)}}=Js^2-\frac{J}{2}(\hat{s}_{i}^{+}\hat{s}_{i+1}^{-}+\hat{s}_{i}^-\hat{s}_{i+1}^+)-J\hat{s}_i^z\hat{s}_{i+1}^z.
\end{align}
where $i=1,2,\cdots,L$ is the site index,  $\hat{s}_i^\pm\coloneqq\hat{s}_i^x\pm i\hat{s}_i^y$, and we assume $J{>}0$. Throughout this work, we assume the periodic boundary condition. 
The Hamiltonian has the $\mathrm{SO}(3)$ spin rotation symmetry generated by $\hat{S}^a\coloneqq\sum_{i=1}^L\hat{s}_i^a$ ($a=x,y,z$) and the time-reversal symmetry. 

Ground states of $\hat{H}^{\text{(FM)}}$ can be expressed as 
\begin{align}
|M\rangle^{\text{(FM)}}\coloneqq \frac{1}{N_M}(\hat{S}^{-})^{sL-M}|\Phi_0\rangle\label{gsferro}
\end{align}
for $M=-sL,-sL+1,\cdots,sL$, where $|\Phi_0\rangle$ is the fully polarized state in the $z$ basis and $N_M>0$ is a normalization factor.  The Hamiltonian  is frustration-free because $(i)$ $\hat{H}_i^{\text{(FM)}}=(J/2)[2s(2s+1)-(\hat{\vec{s}}_i+\hat{\vec{s}}_{i+1})^2]\geq0$ is positive-semidefinite and $(ii)$ $\hat{H}_i^{\text{(FM)}}|M\rangle^{\text{(FM)}}=0$ for every $i$.

\begin{figure}[t]
\includegraphics[width=\columnwidth]{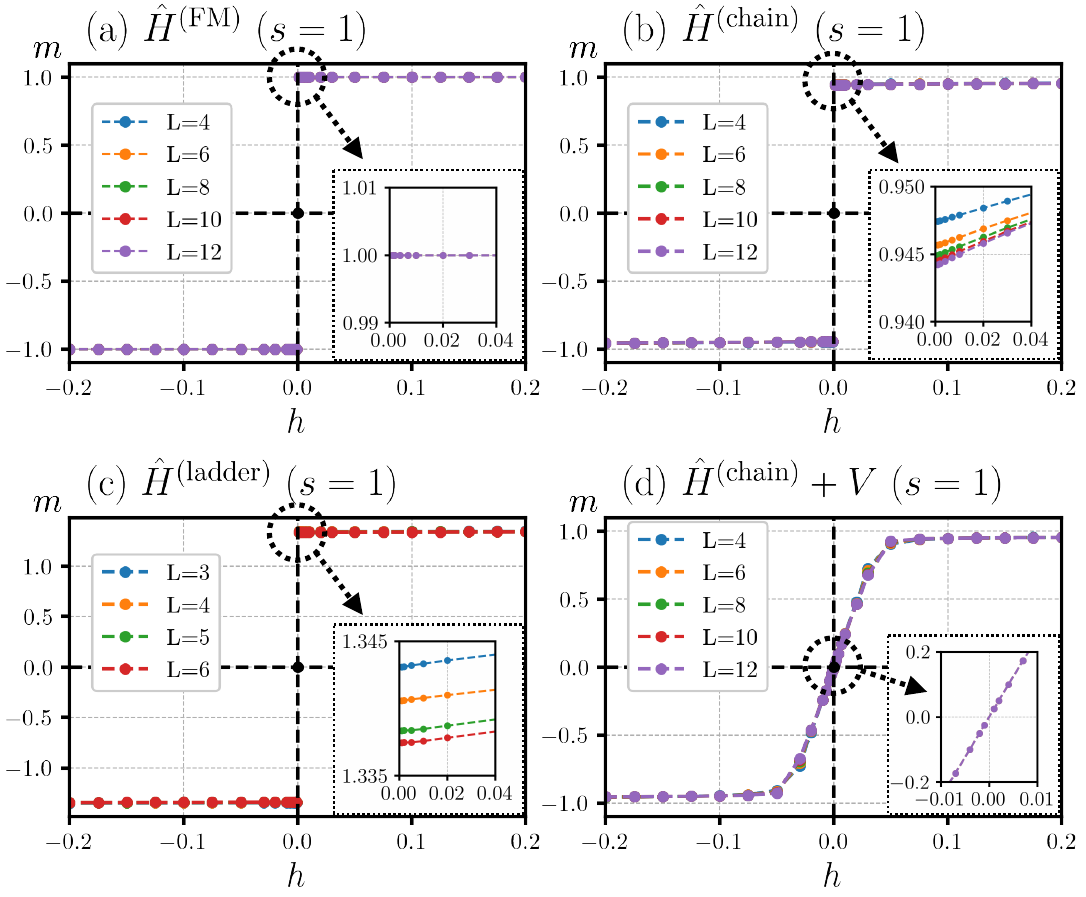}
\caption{\label{fig_order} {\bf Order parameter $m$ as function of symmetry breaking field $h$ for spin-1 systems.} {\bf (a)} Ferromagnetic Heisenberg model Eq.\eqref{eq:FM} with $J\,{=}\,1$. {\bf (b)} Chain model, Eq.\eqref{eq:chain} with $J\,{=}\,1$.
{\bf (c)} Ladder model, Eq.~(\ref{Hamiltonian},\ref{hi}) with $(J_x,J_z,B_z)\,{=}\,(2,1,1)$. {\bf (d)} Chain model with perturbation $\hat{V}$ with $\delta=0.02$. 
The order parameter is $\hat{\mathcal{O}}\,{\coloneqq}\,\hat{S}^z$ for (a), $\hat{\mathcal{O}}\,{\coloneqq}\,\hat{S}^x$ for (b,d), while $\hat{\mathcal{O}}\,{\coloneqq}\,\sum_i \hat{n}^{(1)}_i$ in Eq.\eqref{eq:orderparam} for (c). For a fair comparison, the order parameter in (c) needs to be rescaled by $1/\sqrt{2}$. The insets of (b,c) show the absence of SO(3) symmetry in new models. In (d), the order parameter vanishes as $h$ decreases across all system sizes, demonstrating the absence of U(1) symmetry breaking. The results are obtained by exact diagonalization.}
\end{figure}

The operator $\hat{\mathcal{O}}\,{\coloneqq}\,\hat{S}^z$ can be used as the order parameter for the spontaneous breaking of the symmetries generated by $\hat{Q}=\hat{S}^x$ and $\hat{S}^y$.
We introduce an external field $h$ by adding $-h\hat{\mathcal{O}}$ to the Hamiltonian. 
The external field splits the degeneracy and the ground state becomes unique.  Denoting it by $|\Phi(h)\rangle$, we look at the expectation value
\begin{align}
m(h)\coloneqq\frac{1}{L}\langle \Phi(h)|\hat{\mathcal{O}}|\Phi(h)\rangle, \label{order}
\end{align}
whose magnitude is bounded by $s$.
We show our numerical results in Fig.~\ref{fig_order}(a), where the order parameter saturates the maximum possible value at any finite $h$. This implies the spontaneous breaking of the symmetries generated by $\hat{S}^x$ and $\hat{S}^y$. In general, if the order parameter does not vanish after taking a thermodynamic limit first and then vanishing field limit, i.e., $\lim_{h\to0^+} \lim_{L\to\infty}m(h)\,{\neq}\,0$, the system spontaneously breaks the symmetry.

In the literature~\cite{anderson2018basic,WatanabeMurayamaPRX,BEEKMAN2015461,PhysRevResearch.2.013304}, this symmetry breaking in one dimension has been understood based on the fact that the order parameter $\hat{\mathcal{O}}$ commutes with $\hat{H}^{\text{(FM)}}$. The ground state is a simultaneous eigenstate of $\hat{\mathcal{O}}$ and $\hat{H}^{\text{(FM)}}$ and quantum fluctuations cannot destroy the order.

\emph{Spin Chain Model.---}
The Hamiltonian for the first class of our new examples is given by $\hat{H}^{\text{(chain)}}\coloneqq\sum_{i=1}^L\hat{H}_i^{\text{(chain)}}$ with
\begin{align} \label{eq:chain}
\hat{H}_i^{\text{(chain)}}&= J\big[s(s+1)-(\hat{s}_{i}^z)^2\big]\big[s(s+1)-(\hat{s}_{i+1}^z)^2\big]
\notag\\
&\quad
-\frac{J}{2}(\hat{s}_{i}^{2+}\hat{s}_{i+1}^{2-}+\hat{s}_{i}^{2-}\hat{s}_{i+1}^{2+})-J\hat{s}_{i}^z\hat{s}_{i+1}^z,
\end{align}
where $\hat{s}_i^{2\pm}$ are modified spin-raising/lowering operators defined by
\begin{align}
\hat{s}_i^{2\pm}|m\rangle_i&\coloneqq[s(s+1)-m(m\pm1)]|m\pm1\rangle_i,
\end{align}
which should be compared to the standard spin-raising/lowering operators $\hat{s}_i^{\pm}|m\rangle_i=\sqrt{s(s+1)-m(m\pm1)}|m\,\,{\pm}\,\,1\rangle_i$.
Here, $|m\rangle_i$ is the state at the $i$th site satisfying $\hat{s}_i^z|m\rangle_i\,\,{=}\,\,m|m\rangle_i$.
For $s\,{=}\,1/2$, $\hat{s}^{2\pm}$ coincides with $\hat{s}^{\pm}$ and $\hat{H}^{\text{(chain)}}$ is reduced to $\hat{H}^{\text{(FM)}}$ discussed above.
For $s\,{=}\,1$, $\hat{s}_i^{2\pm}\,{=}\,\sqrt{2}\hat{s}_i^{\pm}$ and the Hamiltonian becomes a modified XXZ model: $\hat{H}_i^{\text{(chain)}}\,{\coloneqq}\,{-}J(2\hat{s}_{i}^x\hat{s}_{i+1}^x+2\hat{s}_{i}^y\hat{s}_{i+1}^y+\hat{s}_{i}^z\hat{s}_{i+1}^z)+J[2\,{-}\,(\hat{s}_i^z)^2][2\,{-}\,(\hat{s}_{i+1}^z)^2]$. For higher spins, the relation between $\hat{s}_i^{2\pm}$ and $\hat{s}_i^{\pm}$ are more complicated, although we can formally write $\hat{s}_i^{2\pm}=\sqrt{s\pm\hat{s}_i^z} \hat{s}_i^\pm \sqrt{s\mp\hat{s}_i^z}$. When $s\,\,{\geq}\,\,1$, the model has the following symmetries: a $\mathrm{U}(1)$ spin rotation about $z$ axis, a $\pi$ spin rotation about $x$ (or $y$) axis, and complex-conjugation~\footnote{Combined with the $\pi$ spin rotation about $y$-axis, it becomes a conventional time-reversal that flips spins.}.  

The ground states of the model are given by the uniform superposition of all orthonormal states in the $\hat{S}^z\,{=}\,M$ sector:
\begin{align}
|M\rangle^{\text{(chain)}}=\frac{1}{N_M'}\sum_{\{m_i\}|\sum_im_i=M}|\{m_i\}\rangle.\label{gschain}
\end{align}
Here, $|\{m_i\}\rangle$ represents the normalized state satisfying $\hat{s}_{i}^z|\{m_i\}\rangle=m_i|\{m_i\}\rangle$, and $N_M'$ is the normalization constant.
In the summation, $\{m_i\}$ runs all spin configurations with ${-}s\,{\leq}\,m_i\,{\leq}\,s$ for all $i$ and $\sum_{i=1}^Lm_i=M$.
One can check easily that $\hat{H}_i^{\text{(chain)}}|M\rangle^{\text{(chain)}}=0$ for every $i$.  Also, as we show later, $\hat{H}_i^{\text{(chain)}}$ is positive semidefinite. Thus, $\hat{H}^{\text{(chain)}}$ is frustration-free. 

For the U(1) symmetry generated by $\hat{Q}\,{\coloneqq}\,\sum_{i=1}^L\hat{S}_i^z$, 
$\hat{\mathcal{O}}\,{\coloneqq}\,\,\hat{S}^x$ can be used as the order parameter. Just as we did for the ferromagnetic case, we apply an external field coupled to $\hat{\mathcal{O}}$ and compute the ground state expectation value in Eq.\eqref{order}. We show our numerical results in Fig.~\ref{fig_order}(b), which clearly displays the spontaneous breaking of the U(1) symmetry generated by $\hat{Q}$. We emphasize that the order parameter $\hat{\mathcal{O}}$ does not commute with the Hamiltonian, which can be inferred from the inset showing that the order parameter does not saturate to the maximum possible value. The value of $\lim_{h\to0^+}m(h)$ is given by the largest eigenvalue of the matrix $(\mathcal{O})_{M,M'}\coloneqq\langle M|\hat{\mathcal{O}}|M'\rangle^{\text{(chain)}}$, as illustrated in Sec. I of the Supplementary Material (SM)\cite{SM}. Interestingly, a marked decrease in the normalized order parameter $m(0^+)/s$ is observed as the spin per site $s$ increases, indicating that there does not exist an SO(3) symmetry emerging at low energies for this model with $s\,{>}\,1/2$.

\emph{Spin Ladder Model.---}
The second class of our new examples is a spin ladder model. 
The Hamiltonian takes the form
\begin{align}  
\hat{H}^{\text{(ladder)}}&\coloneqq\sum_{i=1}^L\hat{H}_i^{\text{(ladder)}} = \sum_{i=1}^L \big(\hat{h}_{i,1}-\hat{h}_{i,2}\big)^2,\label{Hamiltonian}
\end{align}
where $\hat{h}_{i,\alpha}$ describes short-range interactions among spins on the $\alpha$th leg ($\alpha=1,2$), satisfying $\hat{h}^{\vphantom{T}}_{i,2}=\hat{h}_{i,1}^T$. This model is discussed in Ref.~\cite{JongYeonLee} as the effective Hamiltonian of  Brownian random circuits~\cite{Lashkari2013, Gharibyan2018, ZhouXiao2019, XuSwingle2019, XiaoZhou2019}. The model in Eq.\eqref{Hamiltonian} has $(i)$ two U(1) symmetries for each leg and $(ii)$ the anti-unitary $\mathbb{Z}^*_2$ symmetry generated by the leg exchange accompanied by complex conjugation.  

Although the choice of $\hat{h}_{i,\alpha}$ can be quite arbitrary, for concreteness here we assume a typical XXZ Hamiltonian with Zeeman field: 
\begin{align}  
\hat{h}_{i,\alpha}&=J_x\big(\hat{s}_{i,\alpha}^x\hat{s}_{i+1,\alpha}^x+\hat{s}_{i,\alpha}^y\hat{s}_{i+1,\alpha}^y\big)+J_z\hat{s}_{i,\alpha}^z\hat{s}_{i+1,\alpha}^z\notag\\
&\quad+\frac{B_z}{2}\big(\hat{s}_{i,\alpha}^z+\hat{s}_{i+1,\alpha}^z\big)\label{hi},
\end{align}
where two U(1) symmetries are generated by $\hat{S}_\alpha^z\,{\coloneqq}\,\sum_i \hat{S}_{i,\alpha}^z$ for $\alpha\,{=}\,1,2$. This model has an additional symmetry which is the bare complex-conjugation.


The groundstates of Eq.\eqref{Hamiltonian} can be written as~\cite{JongYeonLee}
\begin{align}
|M\rangle^{\text{(ladder)}}=\frac{1}{N_M'}\sum_{\{m_i\}|\sum_im_i=M}|\{m_i\}\rangle_1|\{m_i\}\rangle_2\label{gsladder}
\end{align}
for $M\,{=}\,-sL,-sL+1,\cdots,sL$, 
where $|\{m_i\}\rangle_\alpha$ represents the state satisfying $\hat{s}_{i,\alpha}^z|\{m_i\}\rangle_\alpha\,{=}\,m_i|\{m_i\}\rangle_\alpha$ for the $\alpha$th leg.
Using the relation $\hat{h}^{\vphantom{T}}_{i,2}\,{=}\,\hat{h}_{i,1}^T$, we find $(\hat{h}_{i,1}-\hat{h}_{i,2})|M\rangle^{\text{(ladder)}}\,{=}\,0$ for every $i$. Since the square of a Hermitian operator is positive semidefinite, $\hat{H}^{\text{(ladder)}}$ is frustration-free.
There is no other groundstate as far as $\hat{h}_{i,\alpha}$ is sufficiently general, e.g., Eq.\eqref{hi} with $J_x\neq0$ and $B_z\neq0$.

To diagnose the spontaneous breaking of the diagonal $\mathrm{U}(1)$ symmetry, let us introduce nematic-type order parameters 
\begin{align} \label{eq:orderparam}
    \hat{n}_i^{(1)}& \coloneqq\frac{1}{2} (s_{i,1}^+ s_{i,2}^+ + s_{i,1}^- s_{i,2}^-)=\hat{s}_{i,1}^x\hat{s}_{i,2}^x-\hat{s}_{i,1}^y\hat{s}_{i,2}^y,  \\
    \hat{n}_i^{(2)}& \coloneqq\frac{1}{2i} (s_{i,1}^+ s_{i,2}^+ - s_{i,1}^- s_{i,2}^-)= \hat{s}_{i,1}^x\hat{s}_{i,2}^y+\hat{s}_{i,1}^y\hat{s}_{i,2}^x.
\end{align}
Under the $\mathrm{U}(1)$ symmetry generated by $\hat{Q}\coloneqq\sum_{\alpha=1,2}\hat{S}_\alpha^z$,  they transform as
\begin{align}
&e^{i\theta \hat{Q}}
\begin{pmatrix}
\hat{n}_i^{(1)}\\
\hat{n}_i^{(2)}
\end{pmatrix}
e^{-i\theta \hat{Q}}=
\begin{pmatrix}
\cos2\theta&-\sin2\theta\\
\sin2\theta&\cos2\theta
\end{pmatrix}
\begin{pmatrix}
\hat{n}_i^{(1)}\\
\hat{n}_i^{(2)}
\end{pmatrix}.\label{rotation}
\end{align}
We set $\hat{\mathcal{O}}\coloneqq\sum_{i=1}^L\hat{n}_i^{(1)}$ as the order parameter. Again, the order parameter does not commute with the Hamiltonian, but the plot in Fig.~\ref{fig_order}(c) implies the spontaneous breaking of the diagonal $\mathrm{U}(1)$ symmetry.

\emph{Perturbation.---}To elucidate the mechanism of the spontaneous symmetry breaking in proposed examples, we add the following perturbation to $\hat{H}^{\text{(chain)}}$:
\begin{align}
\hat{V}^{\textrm{(chain)}}&= - \delta \sum_{i=1}^L \hat{s}^z_{i}\cdot\hat{s}^z_{i+1}, 
\label{perturbation}
\end{align}
which preserves the symmetries of $\hat{H}^{\text{(chain)}}$. 

In Fig.~\ref{fig_order}(d), we plot the order parameter when $\delta\,{>}\,0$, showing that $\lim_{h\to0^+}m(h)\,{=}\,0$ and the $\mathrm{U}(1)$ symmetry is unbroken. This observation aligns with the identified gap when $\delta\,{>}\,0$. When $\delta\,{<}\,0$, the system stays gapless, but its dispersion changes from a quadratic to linear form without long-range ordering~(Sec.II of SM). The perturbed Ladder model, which preserves all symmetries yet breaks frustration-freeness, also exhibits an identical behavior.
Interestingly, this behavior mirrors that of perturbed ferromagnets, alluding to a potential link between the Chain and Ferromagnet models. 
More importantly, at any value of $\delta\,{\neq}\,0$, the perturbed Chain Hamiltonian is no longer frustration-free, which suggests that the frustration-free nature plays a critical role in symmetry breaking here. We elaborate on these two aspects in the following sections.

\begin{table*}[t]
\caption{Summary of illustrative examples.}
\label{table}
\centering
\begin{tabular}{c|cccccc}
\hline\hline
Examples							&Generators $\hat{Q}$						&Order Parameter $\hat{\mathcal{O}}$		&Symmetry&$[\hat{H},\hat{\mathcal{O}}]$			&Frustration-Free	&Anderson Tower\\\hline
$\hat{H}^{\text{(FM)}}$				&$\sum_{i=1}^{L}\hat{s}_i^a$ ($a=x,y$)			&$\sum_{i=1}^{L}\hat{s}_i^z$				&Broken&$=0$						&\checkmark				&Absent	\\
$\hat{H}^{\text{(chain)}}$ ($s\geq 1$) 	&$\sum_{i=1}^{L}\hat{s}_{i,1}^z$				&$\sum_{i=1}^{L}\hat{s}_i^x$				&Broken&$\neq0$						&\checkmark				&Absent	\\
$\hat{H}^{\text{(ladder)}}$ 				&$\sum_{i=1}^{L}(\hat{s}_{i,1}^z+\hat{s}_{i,2}^z)$	&$\sum_{i=1}^{L}(\hat{s}_{i,1}^x\hat{s}_{i,2}^x-\hat{s}_{i,1}^y\hat{s}_{i,2}^y)$		&Broken&$\neq0$		&\checkmark	&Absent	\\
$\hat{H}^{\text{(ladder)}}+\hat{V}$ 		&$\sum_{i=1}^{L}(\hat{s}_{i,1}^z+\hat{s}_{i,2}^z)$	&$\sum_{i=1}^{L}(\hat{s}_{i,1}^x\hat{s}_{i,2}^x-\hat{s}_{i,1}^y\hat{s}_{i,2}^y)$		&Unbroken&$\neq0$		& ---	& ---
\\\hline\hline
\end{tabular}
\end{table*}

\emph{Relations among three models.---} 
The three models discussed above have several similarities: $(i)$ they are frustration-free. $(ii)$ there is one ground state in each $\hat{S}^z\,{=}\,M$ sector and the ground state degeneracy is $2sL+1$-fold. $(iii)$ Anderson's tower of states~\cite{anderson2018basic}, which commonly appear in systems with continuous symmetry breaking, is absent and the large $L$ limit is unnecessary to obtain nonzero $\lim_{h\to0^+}m(h)$. See Table~\ref{table} for the comparison of these examples. Here we explain the relation among these models.

First, $\hat{H}^{\text{(chain)}}$ can be obtained from $\hat{H}^{\text{(ladder)}}$ by projecting the $(2s+1)^{2L}$ Hilbert space down to the $(2s+1)^L$ dimensional subspace spanned by
\begin{align}
|\{m_i\}\rangle=|\{m_i\}\rangle_1|\{m_i\}\rangle_2.
\end{align}
We denote this projection by $\hat{\mathcal{P}}$. This relation explains the similarity between the ground states in Eqs.~\eqref{gschain} and~\eqref{gsladder}. Also, the relation $\hat{H}_i^{\text{(chain)}}=\hat{\mathcal{P}}\hat{H}_i^{\text{(ladder)}}\hat{\mathcal{P}}$ proves the positive-semidefinite property of $\hat{H}_i^{\text{(chain)}}$. The coupling constant $J$ in $\hat{H}^{\text{(chain)}}$ is solely determined by $J_x$ in $\hat{H}^{\text{(ladder)}}$ as $J=J_x^2$ and does not depend on $J_z$ or $B_z$.

The relation between $\hat{H}^{\text{(FM)}}$ and $\hat{H}^{\text{(chain)}}$ is more subtle. 
When $s\,{=}\,1/2$, $\hat{H}^{\text{(chain)}}\,{=}\,\hat{H}^{\text{(FM)}}$ as explained above. 
Although $\hat{H}^{\text{(ladder)}}$ has only $\mathrm{U}(1)\,{\times}\,\mathrm{U}(1)$ symmetry, an exact $\mathrm{SO}(3)$ spin rotation symmetry as well as the time-reversal symmetry emerge in the low-energy subspace specified by the projection $\hat{\mathcal{P}}$~\cite{Friedman,Gharibyan2018,Moudgalya2021}.
Consequently, the low-energy effective Lagrangian for $\hat{H}^{\text{(ladder)}}$ should be the same as the one for the ferromagnetic Heisenberg model~\cite{WatanabeMurayamaPRL}.

When $s\,{=}\,1$, $\hat{H}_i^{\text{(chain)}}$ and $\hat{H}_i^{\text{(FM)}}$ can be interpolated by the following one-parameter family 
\begin{align}
\hat{H}_i(\Delta)&\coloneqq-\frac{J}{\Delta}(\hat{s}_{i}^x\hat{s}_{i+1}^x+\hat{s}_{i}^y\hat{s}_{i+1}^y+\Delta\hat{s}_{i}^z\hat{s}_{i+1}^z)\notag\\
&\quad+\frac{J}{\Delta^2}[1-(1-\Delta)(\hat{s}_i^z)^2][1-(1-\Delta)(\hat{s}_{i+1}^z)^2],
\end{align}
where $\Delta\,{=}\,1$ and $\Delta\,{=}\,1/2$ correspond to $\hat{H}_i^{\text{(FM)}}$ and $\hat{H}_i^{\text{(chain)}}$, respectively. 
As far as $\Delta\,{>}\,0$, $\hat{H}(\Delta)=\sum_{i=1}^{L}\hat{H}_i(\Delta)$ is frustration-free and the ground state degeneracy does not depend on $\Delta$.
This interpolation can be obtained by Witten's conjugation~\cite{WITTEN1982, Wouters2021}, and we constructed interpolating Hamiltonian for $s\,\,{\leq}\,\,3$ in Sec.IV of SM.

This frustration-free interpolation unveils a hidden $\mathrm{SO}(3)$ structure within the groundstate manifold of $\hat{H}_i^{\text{(chain)}}$ for $s\,\,{\geq}\,\,1$. To elucidate, we introduce an invertible operator $\hat{\mathcal{M}}=\bigotimes_{i=1}^L\hat{\mathcal{M}}_i$ by $\hat{\mathcal{M}}_i|m\rangle_i\,{=}\,\sqrt{\binom{2s}{s+m} }|m\rangle_i$. 
We find that $\hat{\mathcal{S}}^\pm\coloneqq\hat{\mathcal{M}}\hat{S}^\pm\hat{\mathcal{M}}^{-1}$ connect degenerate ground states of $\hat{H}_i^{\text{(chain)}}$ and satisfy the standard commutation relation of spin-raising/lowering operators, i.e., 
\begin{align}
\hat{\mathcal{S}}^\pm|M\rangle^{\text{(chain)}}\propto |M\pm1\rangle^{\text{(chain)}},\quad[\hat{\mathcal{S}}^+,\hat{\mathcal{S}}^-]=2\hat{S}^z,
\end{align}
similar to the $\hat{S}^\pm$ operators for $\hat{H}_i^{\text{(FM)}}$. See Sec.\,III of SM for the derivation. 
However, it is crucial that $\hat{\mathcal{M}}$ is \emph{non-unitary}, and thus, $[\hat{\mathcal{S}}^\pm,\hat{H}_i^{\text{(chain)}}]\,{\neq}\,0$ and $(\hat{\mathcal{S}}^\pm)^\dagger\,{\neq}\,\hat{\mathcal{S}}^\mp$. 
Therefore, the chain Hamiltonian emerges as an intriguing system characterized by the spontaneous breaking of abelian continuous symmetry in one dimension. In fact, through Witten's conjugation, we can create an infinite family of Hamiltonians with spontaneously broken internal U(1) symmetry starting from the ferromagnetic Hamiltonian.


In summary, we have the following relations among the three models:
\begin{align}
\hat{H}^{\text{(ladder)}}\xrightarrow{\text{Projection $\hat{\mathcal{P}}$}}\hat{H}^{\text{(chain)}}\xrightarrow{\text{Conjugation by $\hat{\mathcal{M}}$}}\hat{H}^{\text{(FM)}}.
\end{align}
Note that the first arrow connects the \emph{low-energy} physics between two, and the second arrow connects the \emph{groundstate manifold} between the two.

\emph{Hohenberg--Mermin--Wagner theorem.---}
Having seen that Abelian continuous symmetries may be spontaneously broken even in one dimension at zero temperature, let us confirm that this result does not contradict any existing no-go theorems. The HMW theorem for finite temperature was originally derived based on the Bogoliubov inequality which is nothing but the Cauchy--Schwarz inequality for a correlation function~\cite{Hohenberg,PhysRevLett.17.1133}.  
This approach was extended to $T\,{=}\,0$ in Ref.~\cite{10.1143/PTP.54.1039,Shastry_1992}.  There is also a proof of Nambu--Goldstone theorem from this direction~\cite{Wagner,Stringari}. Here we reproduce these results with some generalizations.

Let us consider a system defined on a $d$-dimensional lattice $\Lambda$ whose Hamiltonian is given by $\hat{H}\coloneqq\sum_{\bm{i}\in\Lambda}\hat{H}_{\bm{i}}$.
Suppose that a continuous symmetry generated by $\hat{Q}\coloneqq\sum_{\bm{i}\in\Lambda}\hat{Q}_{\bm{i}}$ is spontaneously broken and a Hermitian operator $\hat{\mathcal{O}}\coloneqq\sum_{\bm{i}\in\Lambda}\hat{\mathcal{O}}_{\bm{i}}$ plays the role of the order parameter. We take a Hermitian operator $\hat{X}\coloneqq\sum_{\bm{i}\in\Lambda}\hat{X}_{\bm{i}}$ such that $\hat{\mathcal{O}}\,{=}\,[i\hat{Q},\hat{X}]$.
For example, $\hat{X}_i\,{=}\,\hat{s}_i^y$ and $\hat{\cal O}_i\,{=}\,\hat{s}_i^x$ for $\hat{H}^{\text{(chain)}}$; $\hat{X}_i\,{=}\,(1/2) \hat{n}_i^{(2)}$ and $\hat{\cal O}_i\,{=}\,\hat{n}_i^{(1)}$ for $\hat{H}^{\text{(ladder)}}$.

We start from the $T\,{>}\,0$ version of the theorem. In order to capture long-wavelength fluctuations, we introduce the Fourier transform as $\hat{X}_{\bm{k}}\coloneqq\sum_{\bm{i}\in\Lambda}\hat{X}_{\bm{i}}e^{i\bm{k}\cdot\bm{i}}$ and $\hat{Q}_{\bm{k}}\coloneqq\sum_{\bm{i}\in\Lambda}\hat{Q}_{\bm{i}}e^{i\bm{k}\cdot\bm{i}}$. The Bogoliubov inequality leads to
\begin{align}
\frac{1}{V^2}\sum_{\bm{k}}\langle\hat{X}_{\bm{k}}^\dagger\hat{X}_{\bm{k}}+\hat{X}_{\bm{k}}\hat{X}_{\bm{k}}^\dagger\rangle\geq\frac{1}{V^2}\sum_{\bm{k}}\frac{2T\big|\langle[i\hat{Q}_{\bm{k}}^\dagger,\hat{X}_{\bm{k}}]\rangle\big|^2}{\langle[\hat{Q}_{\bm{k}},[\hat{H}(h),\hat{Q}_{\bm{k}}^\dagger]]\rangle}\label{BI}
\end{align}
for the Gibbs state of $\hat{H}(h)\coloneqq\hat{H}-h\hat{\mathcal{O}}$, where $V$ is the volume of the system. The left-hand side can be written as 
$2\langle\sum_{\bm{i}\in\Lambda}\hat{X}_{\bm{i}}^2\rangle/V$, which remains $O(1)$ even in the limit  {$V\to\infty$} and $h\,{\to}\,0^+$, giving a finite upper bound for the RHS.

On the right-hand side, $\langle[i\hat{Q}_{\bm{k}}^\dagger,\hat{X}^{\vphantom{\dagger}}_{\bm{k}}]\rangle/V$ becomes the expectation value of the order parameter $m(h)\,{:=}\,\langle\hat{\cal O}\rangle/V$ in the $|\bm{k}|\,{\to}\,0$ limit~\footnote{In the lattice model where 
$[i \hat{Q}_{\bm{i}}, \hat{X}_{{\bm{i}}'}] = \hat{\cal O}_r \delta_{{\bm{i}},{\bm{i}}'}$,
 the commutator $\langle[i\hat{Q}_{\bm{k}}^\dagger,\hat{X}^{\vphantom{\dagger}}_{\bm{k}}]\rangle$ is \emph{exactly} $\hat{\cal O}$ for all $\bm{k}$. }.  
Thus, if the symmetry is spontaneously broken ($\lim_{h{\to}0^+} {\lim_{V\to\infty}} m(h) \neq 0$), the numerator of the RHS stays finite as $|\bm{k}|,h\,{\to}\,0^+$.
However, the denominator  {vanishes} as $|\bm{k}|,h\,{\to}\,0^+$ since $\hat{Q}$ is a symmetry generator of $\hat{H}$. Generically, we can write
\begin{align}
& {\lim_{V\to\infty}}\frac{1}{V}\langle[\hat{Q}_{\bm{k}},[\hat{H}(h),\hat{Q}_{\bm{k}}^\dagger]]\rangle=A_{\bm{k}}+hB_{\bm{k}},
\end{align}
where
\begin{align} \label{eq:commutator}
&A_{\bm{k}}\coloneqq {\lim_{h{\to}0^+}\lim_{V\to\infty}}\frac{-1}{V}\sum_{\bm{i},\bm{j}\in\Lambda}\langle[\hat{Q}_{\bm{i}},[\hat{H},\hat{Q}_{\bm{j}}]]\rangle(1{-}\cos[\bm{k}\cdot(\bm{i}-\bm{j})]),\nonumber \\
&B_{\bm{k}}\coloneqq {\lim_{h{\to}0^+}\lim_{V\to\infty}}\frac{-1}{V}\sum_{\bm{i},\bm{j}\in\Lambda}\langle[\hat{Q}_{\bm{i}},[\hat{\mathcal{O}},\hat{Q}_{\bm{j}}]]\rangle\cos[\bm{k}\cdot(\bm{i}-\bm{j})].
\end{align}
For example, $A_k\,{=}\, {\lim_{h{\to}0^+}\lim_{L\to\infty}}(J/L)\sum_{i=1}^L\langle\hat{s}_{i}^{2+}\hat{s}_{i+1}^{2-}+\text{h.c.}\rangle(1-\cos k)$ and 
$B_{k}\,{=}\,m(0^+)$
 for $\hat{H}^{\text{(chain)}}$.
Converting the summation into integral over the range $|k_i|\leq\pi$ and taking the $h\,{\to}\,0^+$ limit, we find
\begin{align} \label{eq:inequality}
 {\lim_{V\to\infty}}\frac{2}{V}\Big\langle\sum_{\bm{i}\in\Lambda}\hat{X}_{\bm{i}}^2\Big\rangle\geq\int\frac{d^dk}{(2\pi)^d}\frac{2T\big|m(0^+)\big|^2}{A_{\bm{k}}}.
\end{align}
Consider a leading order expansion of $A_{\bm{k}}$ such that $|A_{\bm{k}}|\,{\sim}\,C  |\bm{k}|^{2n_0}$ for small $|\bm{k}|$ for some constant $C$. 
Generically, $n_0\,{=}\,1$ and the integral suffers from infrared divergence in $d\,{\leq}\,2$. In order to satisfy the inequality, $m(0^+)$ has to vanish, implying the absence of continuous symmetry breaking at $d\,{\leq}\,2$ for finite temperature~\cite{Hohenberg,PhysRevLett.17.1133}.
In the presence of dipole symmetries, one may have $n_0\,{>}\,1$~\cite{JongYeonLee}. In such a case, the condition for continuous symmetry breaking is modified to $d\,{>}\,2n_0$~\cite{PhysRevB.105.155107,Kapustin}.

Now, the $T\,{=}\,0$ version differs from the above by two points: $(i)$ $2T$ in the Bogoliubov inequality is replaced with the lowest excitation energy $\omega_{\bm{k}}(h)$ of the momentum $\bm{k}$ sector (see Sec.\,V of SM), and $(ii)$ the expectation value is taken with respect to the unique ground state $|\Phi(h)\rangle$ of $\hat{H}(h)$. Here we assume $\omega_{\bm{k}}(0)\,{<}\,v|\bm{k}|^n$ for some constant $v$, where $n\,\,{\geq}\,\,n_0$ due to the Nambu-Goldstone theorem (Sec.\,VI of SM). 
Then, the RHS of Eq.\eqref{eq:inequality} converges when 
\begin{align}
d>2n_0-n.
\end{align}
Normally, $n_0\,{=}\,n\,{=}\,1$ (e.g. superfluid), and continuous symmetry breaking is possible only when $d\,{>}\,1$~\cite{10.1143/PTP.54.1039}.
However, if $2n_0\,{-}\,n\,{<}\,1$, a continuous symmetry breaking is allowed even in $d\,{=}\,1$. In fact, our examples precisely satisfy this condition by $2n_0\,{=}\,n\,{=}\,2$ thanks to the frustration-free property of the Hamiltonian.

\emph{Gapless Excitations in Frustration-Free Systems.---}
There are several recent general results on low-energy excitations in gapless frustration-free systems.
Suppose that $\hat{H}\,{=}\,\sum_{i=1}^L\hat{H}_i$ is frustration-free and translation invariant with zero groundstate energy.
Let us take a length $\ell$ ($3\leq\ell< L/2$) and define subsystem Hamiltonian by $\hat{H}_{\ell,x_0}^{\text{OBC}}\,{\coloneqq}\,\sum_{i=0}^{\ell-2}\hat{H}_{x_0+i}$.
We write the smallest nonzero eigenvalue of $\hat{H}$ and $\hat{H}_{\ell,x_0}^{\text{OBC}}$ as $\epsilon_L^{\text{PBC}}$ and $\epsilon_\ell^{\text{OBC}}$, respectively. 
If $\hat{H}$ is gapless in the sense that $\lim_{L\to\infty}\epsilon_L^{\text{PBC}}\,{=}\,0$, then there exists a constant $C\,{>}\,0$ such that
$0\,{<}\,\epsilon_\ell^{\text{OBC}}\,{<}\,C(\ell^2+\ell)^{-1}$ \cite{Knabe,GossetMozgunov,Anshu}.  
For readers' convenience, we sketch the proof of these results in Sec.~VII of SM.
Therefore, assuming there is no weak edge zero mode~\cite{Verresen}, a spontaneously broken continuous symmetry for the frustration-free Hamiltonian is always accompanied by an excitation whose dispersion is quadratic or softer ($n\,{\geq}\,2$).
This is why spontaneous breaking of continuous symmetries is not prohibited for frustration-free Hamiltonians even in one dimension.


\emph{Conclusion and Outlook.---}
In this Letter, we discussed illuminating spin models in which a $\mathrm{U}(1)$ symmetry is spontaneously broken, where the spontaneous symmetry breaking implies the divergence of the uniform charge susceptibility~\cite{Momoi,Kapustin}. Unlike the case of the Heisenberg ferromagnet, the order parameter does not commute with the Hamiltonian.
We remark that in our models, symmetry generators are written by an unweighted sum of local operators unlike the orientational order in 2d at finite temperature~\cite{Halperin2019} or dipolar symmetry breaking in 1d at zero temperature~\cite{PhysRevB.105.155107}, whose mechanism behind of symmetry breaking is conceptually different.
Furthermore, we have identified the origin of interesting features shared among these models, which is the hidden $\mathrm{SO}(3)$ symmetry realized in a non-unitary manner. This hints at a deeper connection between proposed models with the SSB of abelian U(1) symmetry and the Heisenberg ferromagnet with the SSB of nonabelian SO(3) symmetry breaking. Beyond proposing concrete models, our discussion clarifies that continuous symmetries are allowed to be broken in one dimension as far as the Hamiltonian is frustration-free. 
The field theoretic understanding of these exotic behaviors would be an exciting future direction.

\begin{acknowledgments}
We thank Yohei Fuji, Zijian Xiong, Masaki Oshikawa, Olumakinde Ogunnaike, Johannes Feldmeier, and Huan-Qiang Zhou for useful discussions.
The work of H.W. is supported by JSPS KAKENHI Grant No.~JP20H01825 and JP21H01789.
The work of H.K. is supported by JSPS KAKENHI Grand No.~JP23H01093, No.~JP23H01093, and MEXT KAKENHI Grant-in-Aid for Transformative Research Areas A “Extreme Universe” (KAKENHI Grant No. JP21H05191). The work of J.Y.L is supported by a Simons investigator fund and a faculty startup grant at the University of Illinois, Urbana-Champaign. This work was initiated in part at the Aspen Center for Physics, which is supported by National Science Foundation grant PHY-2210452.
\end{acknowledgments}

\newpage

\onecolumngrid


\makeatletter
\def\l@subsection#1#2{}
\def\l@subsubsection#1#2{}
\makeatother

\setcounter{equation}{0}
\setcounter{figure}{0}
\setcounter{table}{0}

\makeatletter
\renewcommand{\theequation}{S\arabic{equation}}
\renewcommand{\thefigure}{S\arabic{figure}}
\setcounter{subsection}{0}

\begin{center}
    \textbf{\large Supplementary Material for \\ \vspace{7pt}
    ``Spontaneous breaking of U(1) symmetry at zero temperature in one dimension''}

    \vskip6mm
    
    {\noindent \normalsize  Haruki Watanabe, Hosho Katsura, and Jong Yeon Lee}

    \vskip3mm

    \end{center}

\vspace{10pt}
    
In this supplementary material, we elaborate on several technical details that support the claim made in the manuscript. In \secref{sec:order_param} and \secref{sec:perturbed}, we provide exact diagonalization results of the order parameters and excitation spectra for the (perturbed) chain model with local spins $s=1/2,1,3/2,2$. In \secref{sec:SO3}, we elaborate on the mathematical details underlying a hidden SO(3) symmetry realized in a non-unitary manner. In \secref{sec:inter}, we provide technical details to construct a U(1)-symmetric Hamiltonian with spontaneous symmetry breaking interpolating the ferromagnetic Heisenberg and the proposed Chain models. In \secref{sec:HMW} and \secref{sec:NG}, we provide concise derivations of the zero-temperature Hohenberg-Mermin-Wagner theorem and Nambu-Goldstone theorem using the Bogoliubov inequality. Finally, in \secref{sec:FF}, we review theorems on frustration-free Hamiltonians.

\tableofcontents

\section{Order parameter in the chain model} \label{sec:order_param}
    
As discussed in the main text, the groundstate manifold of the chain Hamiltonian is $2sL+1$-fold degenerate. Each groundstate is denoted as 
\begin{align} \label{eq:gs}
    |M\rangle^{\text{(chain)}}=\frac{1}{ \sqrt{ N_M' } }\sum_{\{m_i\}|\sum_im_i=M}|\{m_i\}\rangle,
\end{align}
which is the uniform superposition of all configurations with a net magnetization in $z$-direction $M$. Here, $N_M'$ is the total number of such configurations.

From the degenerate perturbation theory, the order parameter $\langle \hat{S}^x \rangle$ under the infinitesimal field $-h \hat{S}^x$ can be obtained by finding the largest eigenvalue of the matrix $(\hat{S}^x)_{M,M'} := \langle M| \hat{S}^x |M' \rangle$. Since $\hat{S}^x\,{=}\,\frac{1}{2}\sum_i (\hat{S}_i^+ + \hat{S}_i^-)$, $(\hat{S}^x)_{M,M'}$ is non-zero only if $M'\,{=}\, M \,{\pm}\,1$. In order to proceed, let $\tilde{M}\,{=}\, M\,{+}\,sL \in \{ 0,1,2,...,2sL \}$. Then, the normalization constant $N_M'$ is given as the coefficient of the term $x^{M+sL}$ when we expand the polynomial $(1+x+...+x^{2s})^L$, i.e., 
\begin{align}
    \Big( \sum_{m=0}^{2s} x^m \Big)^L = \sum_{M=-sL}^{sL} N_M' \cdot x^{M+sL} = \sum_{\tilde{M}=0}^{2sL} N_{\tilde{M}-sL}' \cdot x^{\tilde{M}}.
\end{align}
In order to make $s$ and $L$ dependence of the $N_{M}'$ more  {explicit}, let us denote $N_{\tilde{M}-sL}'$ by $f(\tilde{M},L,s)$. Then,
\begin{align}
    f(\tilde{M},L,s) =  \hspace{-15pt} \sum_{  \substack{\{t_n\}, t_n \geq 0 \\ \sum_n t_n = L,\,\, \sum_n n t_n = \tilde{M} }  }  \hspace{-15pt}  \frac{L!}{t_0! t_1! \cdots t_{2s}!}
\end{align}
where $t_n$ is the number of sites with $s^z_i = n-s$. We remark that 
the following relation 
allows one to iteratively find $f$ for larger values of $(\tilde{M},L,s)$:
\begin{align}
    f(\tilde{M},L,s) &= \sum_{m=0}^{2s} f(\tilde{M}-m,L-1,s) \nonumber \\
    f(m,1,s) &= \begin{cases}
        1  \quad & \textrm{if } 0 \leq m \leq 2s \\
        0 \quad & \textrm{otherwise}.
    \end{cases}
\end{align}

Now we are ready to calculate the matrix $\hat{S}^x$ in the groundstate manifold. Using the translation invariance of the groundstates, we can show that 
\begin{align}
        \langle M | \hat{S}^x | M-1 \rangle &= \frac{L}{2} \langle M |  S_1^+ | M-1 \rangle = \frac{L}{2} \frac{1}{\sqrt{N_M N_{M-1}}} \sum_{ \substack{\{m_i \} \\ \sum_i m_i = M} } \sum_{ \substack{\{m'_i \} \\ \sum_i m'_i = M-1} } \langle \{m_i\} | \hat{S}_1^+ | \{m'_i\} \rangle \nonumber \\
    &= \frac{L}{2} \frac{1}{\sqrt{N'_M N'_{M-1}}} \sum_{m=-s+1}^s \sum_{ \substack{\{m_i \}_{i=2} \\ \sum_{i=2} m_i = M-m} } \sum_{ \substack{\{m'_i \}_{i=2} \\ \sum_{i=2} m'_i = M-m} } \langle m | S_1^+ | m-1 \rangle \cdot \langle \{m_i\} |  \{m'_i\} \rangle \nonumber \\
    &= \frac{L}{2} \frac{1}{\sqrt{N'_M N'_{M-1}}} \sum_{m=-s+1}^s  f(\tilde{M}-m, L-1, s) \cdot  \langle m | S_1^+ | m-1 \rangle   \nonumber \\
    &= \frac{L}{2} \frac{1}{\sqrt{N'_M N'_{M-1}}} \sum_{m=-s+1}^s  f(\tilde{M}-m, L-1, s) \cdot \sqrt{s(s+1) - m(m-1)}.
\end{align}
    
\begin{figure}[t]
    \includegraphics[width=1\columnwidth]{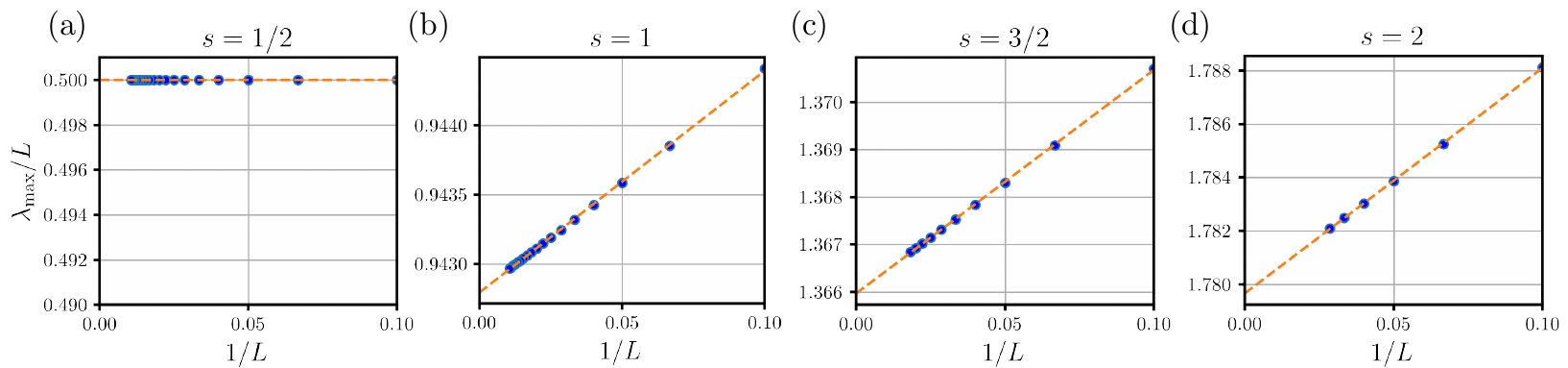}
    \caption{\label{fig:eig} {\bf Normalized maximum eigenvalue of $(\hat{S}^x)_{M,M'}$ as a function of {inverse} system size.} Here, we plot maximum eigenvalues of $\hat{S}^x = \sum_i \hat{S}^x_i$ in the groundstate basis, which are normalized by the system size $L$. Here 
$s$ is the spin per site. Each figure corresponds to the model with (a) $s=1/2$, (b) $s=1$, (c) $s=3/2$, and (d) $s=2$. We observe that the maximum value is smaller than $s$, and as we increase $s$, this deviation increases.  }
\end{figure}

In Fig.~\ref{fig:eig}, we plot its maximum eigenvalue $\lambda_\mathrm{max}$ normalized by the system size $L$ as a function of $1/L$, which corresponds to the order parameter $\lim_{h\to0^+} \lim_{L\to\infty}m(h)$. The result shows that the order parameter does not saturate to its maximum possible value $s$. Furthermore, the discrepancy between the order parameter and $s$ increases as $s$ increases.

\begin{figure}[t]
    \includegraphics[width=\columnwidth]{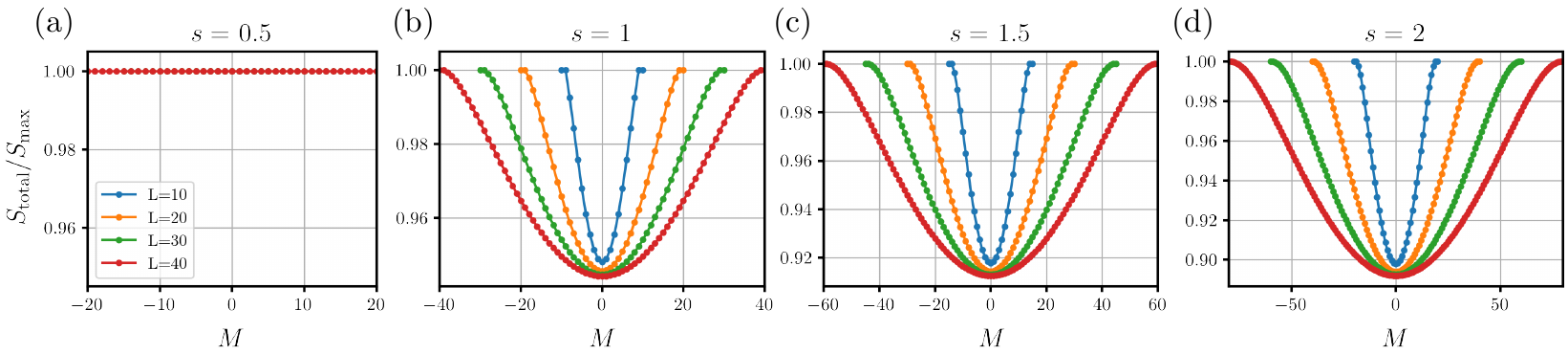}
    \caption{\label{fig:SO3} {\bf Total spin magnitude in the groundstate manifold of the chain Hamiltonian.} Here, we plot the total spin magnitude $S_\mathrm{total}$ normalized by its maximum possible value $S_\mathrm{max} = \sqrt{sL(sL+1)}$ as function of $S^z=M$ that labels the groundstate. Each figure corresponds to the model with (a) $s=1/2$, (b) $s=1$, (c) $s=3/2$, and (d) $s=2$. }
\end{figure}

Observing that the chain model order parameter does not achieve the maximum possible value $s$, we conclude that there does not exist any SO(3) symmetry emerging at low energies.  
This can be also observed by simply examining a total spin magnitude $ {\hat{S}^2} =  {(\hat{S}^x)^2 + (\hat{S}^y)^2 + (\hat{S}^z)^2}$ of Eq.~(\ref{eq:gs}). In the ferromagnetic Heisenberg model, all groundstates have the maximum value $\hat{S}^2 = sL(sL+1)$. However, as illustrated in Fig.~\ref{fig:SO3}, groundstates of the chain Hamiltonian with $s \geq 1$ can have smaller $\hat{S}^2$ values away from $M = \pm sL$. 


    \section{Perturbed chain Model} \label{sec:perturbed}
    
    In this section, we study the spectrum of the perturbed chain model with $s=1$ {and system size $L$}, whose Hamiltonian is given by 
    \begin{align} \label{eq:chain}
        {\hat H} &= \sum_i \hat{H}_i^{\text{(chain)}} + \sum_i \hat{V}_i^{\text{(chain)}}   \nonumber \\
        \hat{H}_i^{\text{(chain)}}&= {-}J(2\hat{s}_{i}^x\hat{s}_{i+1}^x+2\hat{s}_{i}^y\hat{s}_{i+1}^y+\hat{s}_{i}^z\hat{s}_{i+1}^z)+J[2\,{-}\,(\hat{s}_i^z)^2][2\,{-}\,(\hat{s}_{i+1}^z)^2], \nonumber \\
        \hat{V}^{\textrm{(chain)}}_i&= - \delta  \hat{s}^z_{i}\cdot\hat{s}^z_{i+1}.
    \end{align} 
     {In the following, we assume $J>0$ and impose periodic boundary conditions.}
    Before getting into the details, we first remark that when $L$ is even, {${\hat H}$ can be written as
    \begin{align}
    {\hat H} = {\hat U}_{{\rm even},\pi} \left[
    2J {\hat H}_{\rm XXZD} (\Delta, D) + J \sum_i  (\hat{s}_i^z)^2 (\hat{s}_{i+1}^z)^2 +{\rm const.}
    \right] {\hat U}_{{\rm even}, \pi},
    \end{align}
    with $\Delta = -\tfrac{J+\delta}{2J}$ and $D=-2$. Here, ${\hat U}_{{\rm even},\pi}$ is a $\pi$ rotation about the $z$-axis on the even sites defined by ${\hat U}_{{\rm even},\pi} = \prod_{j:{\rm even}} \exp (-i \pi {\hat S}^z_j)$ and ${\hat H}_{\rm XXZD} (\Delta, D) $ is the spin-$1$ XXZ Hamiltonian with single-ion anisotropy:
    \begin{align} \label{eq:XXZD}
    {\hat H}_{\rm XXZD} (\Delta, D)  = \sum_i \left(
    \hat{s}_{i}^x\hat{s}_{i+1}^x + \hat{s}_{i}^y\hat{s}_{i+1}^y
    +\Delta \hat{s}_{i}^z \hat{s}_{i+1}^z
    \right)
    +D \sum_i (\hat{s}_i^z)^2,
    \end{align}
    which has been extensively studied in the context of the Haldane conjecture~\cite{botet1983finite, schulz1986phase, den1989preroughening, chen2003ground}. 
    Second, the model Eq.~(\ref{eq:chain}) has  the groundstate degeneracy of at most $2L+1$, which is 
    a consequence of the Perron-Frobenius theorem: 

    \smallskip 

    {\noindent \emph{Proof}}: The Perron-Frobenius theorem implies that if a real square matrix $A$ is $(i)$ nonnegative and $(ii)$ irreducible, it has a unique largest eigenvalue $r$ and the corresponding eigenvector can be written such that its entries are all strictly positive.
    By irreducible, what it means is that any two basis  {vectors} $|i \rangle$ and $|i' \rangle$ are connected by the successive application of the matrix $A$. 
    Now, consider  $H|_{S^z = M}$, which is the Hamiltonian ${\hat H}$ restricted to a $\hat{S}^z = M$ sector. Let us take $A = cI - H|_{S^z = M}$ for some large value of $c$ such that $A$ is nonnegative. 
    In each $S^z$ sector, a pair of two basis vectors can be always connected by the successive applications of $A$ because $\sum_i \hat{s}_i^+ \hat{s}_{i+1}^- + \textrm{h.c.}$ can connect all different basis  {vectors} after applied enough~\cite{TasakiBook}. 
    As $A$ has the unique largest eigenvalue, $H|_{S^z = M}$ has the unique smallest eigenvalue. This holds for a generic value of $s$. Therefore, the chain Hamiltonian ${\hat H}^{({\rm chain})}$ proposed in the main text has the groundstate degeneracy of at most $2sL+1$.  \hfill $\square$.

    \smallskip 

    In Fig.~\ref{fig:dispersion}, we show exact diagonalization results to illustrate the behavior under perturbation where we set $J=1$. In the following, we describe properties of the model in three different regimes of $\delta=0$, $\delta > 0$, and $\delta <0$, which demonstrates its similarity to the spin-1 ferromagnetic Heisenberg model perturbed with an Ising interaction, i.e., $\hat{H}_\mathrm{XXZD}$ in Eq.\eqref{eq:XXZD} with $D=0$ and $\Delta = -1$, $\Delta < -1$ and $\Delta > -1$.

    \begin{figure}[t]
        \includegraphics[width=\columnwidth]{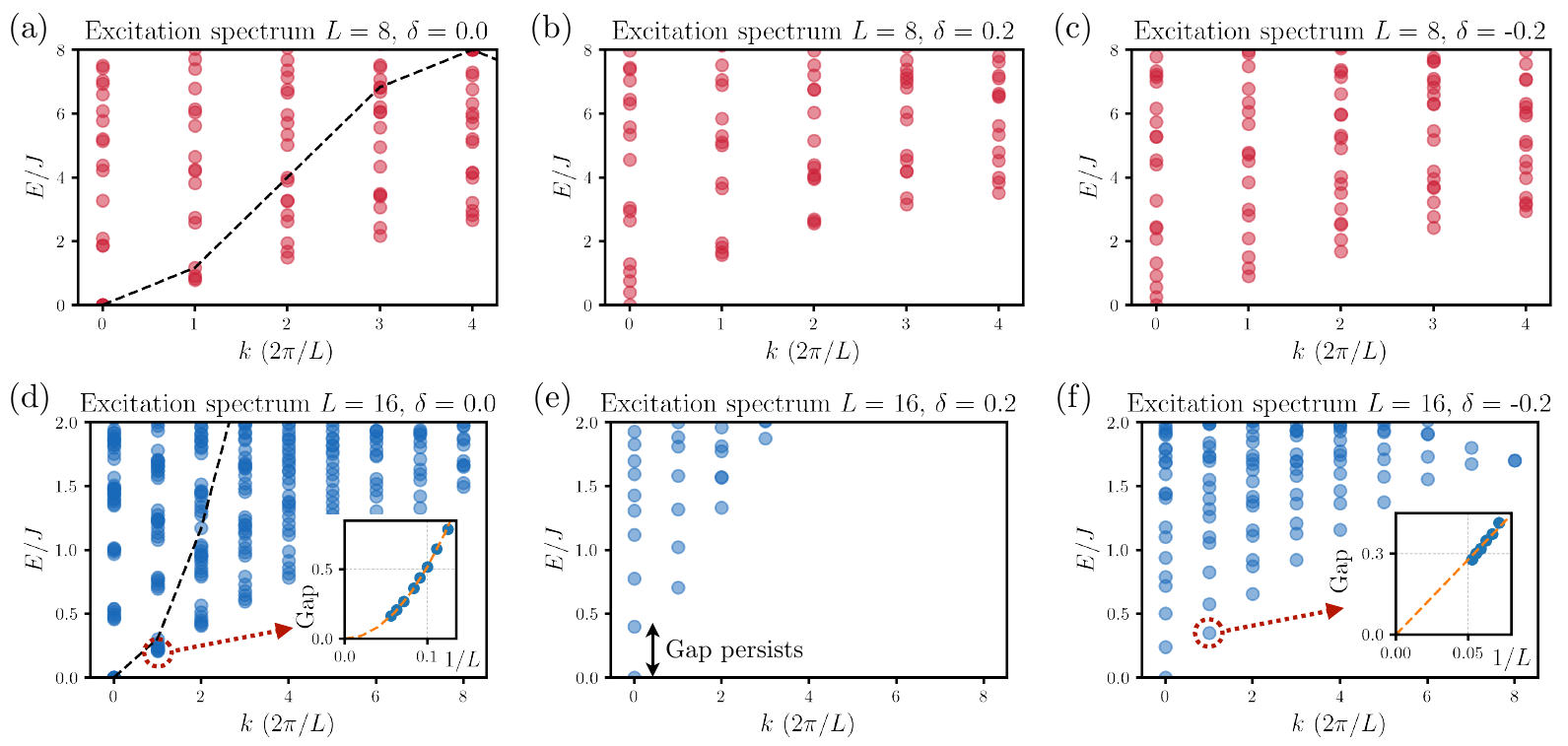}
        \caption{\label{fig:dispersion} {\bf Dispersion for perturbed chain model.} For $\delta \neq 0$, spectrums are shifted so that the groundstate has zero energy. (a) $(L,\delta) = (8,0)$, (b) $(L,\delta) = (8,0.2)$, (c) $(L,\delta) = (8,-0.2)$, (d) $(L,\delta) = (16,0)$, (e) $(L,\delta) = (16,0.2)$, (f) $(L,\delta) = (16,-0.2)$. In (a,d), the groundstate degeneracy at $k=0$ is $2L+1$. The black dashed line in (a,d) corresponds to the analytically solvable dispersion in the $S^z = \pm (L-1)$ sector, $E_k = 4(1-\cos k)$.       
        In (b,e), the groundstate degeneracy is $2$, appearing in the $S^z = \pm L$ sector, and the gap of $2 \delta = 0.4$ persists across all system sizes. This implies the presence of $\mathbb{Z}_2$ symmetry breaking. In (c,f), the groundstate is unique, appearing in the $S^z = 0$ sector. The insets for (d) and (f) plot the gaps between the groundstate and the lowest $k\neq 0$ excitation, along with quadratic ($1/L^2$) and linear ($1/L$) fitting functions, respectively~\footnote{Although the gap between the groundstate and the excitation at $k=0$ is smaller, it is a constant determined by the strength of $|\delta|$. Thus this gap is expected to be irrelevant in the thermodynamic limit.}.  }
    \end{figure}

    \subsection{\texorpdfstring{$\delta = 0$}{delta=0}}

The model under investigation is frustration-free, with its groundstate exhibiting $2L+1$-fold degeneracy. As we elaborated in the main text, there is a zero-energy state for each $S^z$ sector, giving rise to $2L+1$ states that saturate the aforementioned bound from the Perron-Frobenius theorem.

    Furthermore, the model is exactly solvable in the $S^z = \pm (L-1)$ sectors. The black dashed line in Fig.~\ref{fig:dispersion}(a,d) corresponds to this analytically solvable line where $E = 4(1-\cos k)$. 
    This branch at $S^z = \pm (L-1)$ agrees with that of the ferromagnetic Heisenberg model. 
    However, while the ferromagnetic Heisenberg model has exact degeneracy at this point due to SO(3) symmetry, the chain model has SO(3) symmetry \emph{only} at the groundstate manifold, thus there exist some excitations at $k=2\pi/L$ with energies lower than $4(1-\cos k)$. 
    Nevertheless, as we increase the system size, we can observe $1/L^2$ scaling of the gap as illustrated in the inset of Fig.~\ref{fig:dispersion}(d), where the data is obtained for $L \in [8,18]$.

    Note that the plot shows excitations with $k \geq 2$ whose energies are significantly lower than the single particle energy denoted by the dashed line. 
    This is originated from bound states of multiple excitations, which is the generic feature of the quadratically dispersing system. Roughly speaking, two excitations with $k=1$ would have a net energy lower than that of a single particle excitation with $k=2$ due to $E\sim k^2$ dispersion. For example, at $k=2$, the lowest energy of the $S^z = (L-1)$ branch would be much higher than the lowest energy of the $S^z = (L-2)$ branch; this latter branch corresponds to the bound state of two excitations with $k=1$. 

    \subsection{\texorpdfstring{$\delta > 0$}{delta>0}}
    
    The model is no longer frustration-free. 
    Interestingly, the model has the two-fold groundstate degeneracy arising from the $S^z = \pm L$ sectors. The first excited states are coming from $S^z = \pm ({L}-1)$ sectors and the gap is exactly given by $2 \delta$ which persists across all system sizes examined. This behavior is very similar to the ferromagnetic Hamiltonian perturbed by the Ising interaction, which immediately leads to the ferromagnetic ordering in $z$-direction.
    This corresponds to the spontaneous breaking of the $\mathbb{Z}_2$ symmetry generated by a $\pi$ rotation about the $x$ (or $y$)-axis, and the spectrum is gapped.        
    
    \subsection{\texorpdfstring{$\delta < 0$}{delta<0}}    
    
    Again, the model is not frustration-free. The model has a unique groundstate at $S_z = 0$ sector. Furthermore, a careful analysis shows that its gap scales as $1/L$, as shown in the inset of Fig.~\ref{fig:dispersion}(f). This low-energy behavior is captured by the Tomonaga-Luttinger liquid theory, {similar to the XY phases of $H_{\rm XXZD} (\Delta, D)$ in Eq. (\ref{eq:XXZD})~\cite{botet1983finite, schulz1986phase, den1989preroughening, chen2003ground}. }

\section{hidden SO(3) structure} \label{sec:SO3}
As explained in the main text, the groundstates of $\hat{H}^{\text{(FM)}}$ and $\hat{H}^{\text{(chain)}}$ can be expressed as 
\begin{align}
&|M\rangle^{\text{(FM)}}\coloneqq \frac{1}{N_M}(\hat{S}^{-})^{sL-M}|\Phi_0\rangle,\\
&|M\rangle^{\text{(chain)}}=\frac{1}{N_M'}\sum_{\{m_i\}|\sum_im_i=M}|\{m_i\}\rangle.
\end{align}
for $M=-sL,-sL+1,\cdots,sL$. We find
\begin{align} \label{eq:normalizations}
N_M=\sqrt{\frac{(2sL)!(sL-M)!}{(sL+M)!}},\\
N_M'=\sqrt{\sum_{\{m_i\}|\sum_im_i=M}1}.
\end{align}
Here we discuss their relations.

\begin{align}
|M\rangle^{\text{(FM)}}&= \frac{1}{N_M}\sum_{\{k_i\}|\sum_{i=1}^Lk_i=sL-M,0\leq k_i\leq 2s}\frac{(sL-M)!}{\prod_{i=1}^Lk_i!}\prod_{i=1}^L(\hat{s}_i^-)^{k_i}|\Phi_0\rangle\notag\\
&=\frac{(sL-M)!}{N_M}\sum_{\{k_i\}|\sum_{i=1}^Lk_i=sL-M,0\leq k_i\leq 2s}\prod_{i=1}^L\binom{2s}{k_i}^{1/2}|\{m_i=s-k_i\}\rangle\notag\\
&=\frac{(sL-M)!}{N_M}\sum_{\{m_i\}|\sum_im_i=M}\prod_{i=1}^L\binom{2s}{s+m_i}^{1/2}|\{m_i\}\rangle\notag\\
&=\frac{(sL-M)!}{N_M}\sum_{\{m_i\}|\sum_im_i=M}\hat{\mathcal{M}}^{-1}|\{m_i\}\rangle\notag\\
&=\frac{(sL-M)!N_M'}{N_M}\hat{\mathcal{M}}^{-1}|M\rangle^{\text{(chain)}},
\end{align}
where we used $(\hat{s}_i^-)^{k}|s\rangle_i=\sqrt{\frac{(2s)!k!}{(2s-k)!}}|s-k\rangle_i$.

Similarly, starting from the `all down' state $|\Phi_0'\rangle$, we can write the groundstates of  $\hat{H}^{\text{(FM)}}$ as
\begin{align}
&|M\rangle^{\text{(FM)}}= \frac{1}{N_{-M}}(\hat{S}^{+})^{sL+M}|\Phi_0'\rangle=\frac{(sL+M)!N_M'}{N_{-M}}\hat{\mathcal{M}}^{-1}|M\rangle^{\text{(chain)}}.
\end{align}

Therefore,
\begin{align}
\hat{\mathcal{M}}|M\rangle^{\text{(FM)}}=\frac{(sL-M)!N_M'}{N_M}|M\rangle^{\text{(chain)}}=\frac{(sL+M)!N_M'}{N_{-M}}|M\rangle^{\text{(chain)}}.
\end{align}
Since $\hat{\mathcal{M}}|\Phi_0\rangle=|\Phi_0\rangle$ and $\hat{\mathcal{M}}|\Phi_0'\rangle=|\Phi_0'\rangle$, we obtain
\begin{align}
|M\rangle^{\text{(chain)}}=\frac{1}{(sL-M)!N_M'}(\hat{\mathcal{M}}\hat{S}^{-}\hat{\mathcal{M}}^{-1})^{sL-M}|\Phi_0\rangle=\frac{1}{(sL+M)!N_M'}(\hat{\mathcal{M}}\hat{S}^{+}\hat{\mathcal{M}}^{-1})^{sL+M}|\Phi_0'\rangle.
\end{align}

\section{Interpolating Hamiltonian} \label{sec:inter}

In this section, we discuss the interpolating Hamiltonian between the ferromagnetic Heisenberg model
\begin{align}
\hat{H}^{\text{(FM)}}&\coloneqq\sum_{i=1}^L\hat{H}_i^{\text{(FM)}},\\
\hat{H}_i^{\text{(FM)}}&\coloneqq Js^2-\frac{J}{2}(\hat{s}_{i}^{+}\hat{s}_{i+1}^{-}+\hat{s}_{i}^-\hat{s}_{i+1}^+)-J\hat{s}_i^z\hat{s}_{i+1}^z
\end{align}
and our spin chain model
\begin{align} 
\hat{H}^{\text{(chain)}}&\coloneqq\sum_{i=1}^L\hat{H}_i^{\text{(chain)}},\\
\hat{H}_i^{\text{(chain)}}&\coloneqq J\big[s(s+1)-(\hat{s}_{i}^z)^2\big]\big[s(s+1)-(\hat{s}_{i+1}^z)^2\big]-\frac{J}{2}(\hat{s}_i^{2+}\hat{s}_{i+1}^{2-}+\hat{s}_i^{2-}\hat{s}_{i+1}^{2+})-J\hat{s}_{i}^z\hat{s}_{i+1}^z.
\end{align}

Let $|\Phi_0\rangle=\bigotimes_{i=1}^L |s\rangle_i$ be the fully polarized state. 
The coherent state defined by
\begin{align}
|\alpha\rangle^{\text{(FM)}}\coloneqq e^{\alpha\hat{S}^{-}}|\Phi_0\rangle=\bigotimes_{i=1}^L e^{\alpha\hat{s}_i^{-}}|s\rangle_i
\end{align}
is a groundstate of $\hat{H}^{\text{(FM)}}$ regardless of $\alpha\in\mathbb{C}$, because $\hat{S}^{-}\coloneqq \sum_{i=1}^L\hat{s}_i^-$ commutes with $\hat{H}^{\text{(FM)}}$. Indeed, this state can be written as $|M\rangle^{\text{(FM)}}$ as
\begin{align}
|\alpha\rangle^{\text{(FM)}}=\bigotimes_{i=1}^L \sum_{k=0}^{2s}\frac{\alpha^k}{k!}(\hat{s}_i^{-})^k|s\rangle_i=\bigotimes_{i=1}^L \sum_{m_i=-s}^{s}\alpha^{s-m_i}\binom{2s}{s+m_i}^{1/2}|m_i\rangle_i
=\sum_{M=-sL}^{sL}\alpha^{sL-M}\binom{2sL}{sL+M}^{1/2}|M\rangle^{\text{(FM)}}.
\end{align}

On the other hand,
\begin{align}
|\alpha\rangle^{\text{(chain)}}\coloneqq \bigotimes_{i=1}^L \sum_{m_i=-s}^{s}\alpha^{s-m_i}|m_i\rangle_i
=\sum_{M=-sL}^{sL}\alpha^{sL-M}N_M'|M\rangle^{\text{(chain)}}.
\end{align}
is a groundstate of $\hat{H}^{\text{(chain)}}$ regardless of $\alpha\in\mathbb{C}$, because it is given as a superposition of $|M\rangle^{\text{(chain)}}$. The normalization factor $N_M'$ is defined in Eq.\eqref{eq:normalizations}.

To interpolate $|\alpha\rangle^{\text{(FM)}}$ to $|\alpha\rangle^{\text{(chain)}}$, we introduce $q$-bracket, $q$-factorial, and $q$-binomial, respectively, by
\begin{align}
&[m]_q\coloneqq \frac{1-q^m}{1-q}=1+q+\cdots+q^{m-1},\\
&[n]_q!\coloneqq [n]_q[n-1]_q\cdots [2]_q[1]_q,\\
&\binom{n}{k}_q\coloneqq\frac{[n]_q!}{[k]_q![n-k]_q!}.
\end{align}
We introduce an operator $\hat{\mathcal{M}}(q)$ by 
\begin{align}
&\hat{\mathcal{M}}(q)=\bigotimes_{i=1}^L\hat{\mathcal{M}}_i(q),\\
&\hat{\mathcal{M}}_i(q)|m\rangle_i=\binom{2s}{s+m}_q^{-1/2}|m\rangle_i
\end{align}
and define
\begin{align}
|\alpha(q)\rangle\coloneqq\hat{\mathcal{M}}(q)|\alpha\rangle^{\text{(FM)}}.\label{alphaq}
\end{align}

Since the $q$-binomial satisfies
\begin{align}
&\binom{n}{k}_{q=1}=\binom{n}{k},\\
&\binom{n}{k}_{q=0}=1,
\end{align}
$|\alpha(q)\rangle$ coincides with $|\alpha\rangle^{\text{(FM)}}$ at $q=0$ and with $|\alpha\rangle^{\text{(chain)}}$ at $q=1$. 

We postulate the following form of the interpolating Hamiltonian:
\begin{align} 
\hat{H}(q)&\coloneqq\sum_{i=1}^L\hat{H}_i(q),\\
\hat{H}_i(q)&\coloneqq J\Big[\sum_{l=0}^sc_l(q)(\hat{s}_{i}^z)^{2l}\Big]\Big[\sum_{l=0}^sc_l(q)(\hat{s}_{i+1}^z)^{2l}\Big]-\frac{J}{2}(\hat{s}_i^+(q)\hat{s}_{i+1}^-(q)+\text{h.c.})-J\hat{s}_{i}^z\hat{s}_{i+1}^z,
\end{align}
where $\hat{s}_i^-(q)$ is a generalized lowering operator defined by
\begin{align} 
&\hat{s}_i^-(q)|m\rangle_i=w_m(q)|m-1\rangle_i
\end{align}
for $m=-s+1,-s+2,\cdots, s$ and $\hat{s}_i^+(q)\coloneqq(\hat{s}_i^-(q))^\dagger$. Furthermore, we assume $w_{s-(m-1)}(q)=w_{m-s}(q)$ for $1\leq m\leq s$. 
We require that $\hat{H}_i(q)$ to be positive semi-definite. With this condition, one can determine the coefficients $c_0(q), c_1(q),\cdots, c_s(q)$ and $w_{s}(q), w_{s-1}(q), \cdots$ by solving that $|\alpha(q)\rangle$ vanishes under the application of $\hat{H}(q)$ regardless of $\alpha\in\mathbb{C}$. In practice, we can establish a system of equations by imposing $\hat{H}_i(q)$ to vanish for all two-site states obtained by projecting $|\alpha(q)\rangle$ onto each magnetization-$M$ sector.
Note that $\hat{H}_i(q) |\alpha(q) \rangle = 0$ if and only if $\hat{H}_i(q) {\cal P}_M |\alpha(q)\rangle = 0$ for all $M$ due to U(1) symmetry, where ${\cal P}_m$ is a projection onto a sector with $z$-magnetization $M$.

Below we present the solution for $s=1/2, 1, 3/2, 2, 5/2$, and $3$.
\subsubsection{$s=1/2$}
We find
\begin{align}
c_0(q)=\frac{1}{2},
\end{align}
and
\begin{align}
w_{\frac{1}{2}}(q)=1.
\end{align}
The resulting Hamiltonian is the ferromagnetic Heisenberg model:
\begin{align} 
\hat{H}_i(q)= \frac{J}{4}-\frac{J}{2}(\hat{s}_i^+\hat{s}_{i+1}^-+\text{h.c.})-J\hat{s}_{i}^z\hat{s}_{i+1}^z
\end{align}

\subsubsection{$s=1$}
We find
\begin{align}
&c_0(q)=q+1,\\
&c_1(q)=-q,
\end{align}
and
\begin{align}
w_{1}(q)=\sqrt{2} \sqrt{q+1}.
\end{align}

The resulting interpolating Hamiltonian
\begin{align} 
\hat{H}_i(q)&= J\Big[(q+1)-q(\hat{s}_{i}^z)^{2}\Big]\Big[(q+1)-q(\hat{s}_{i+1}^z)^{2}\Big]-\frac{J}{2}(q+1)(\hat{s}_i^+\hat{s}_{i+1}^-+\text{h.c.})-J\hat{s}_{i}^z\hat{s}_{i+1}^z
\end{align}
agrees with  $\hat{H}_i(\Delta)$ in the main text if $\Delta$ is replaced with $1/(q+1)$.

\subsubsection{$s=3/2$}
We find
\begin{align}
&c_0(q)=\frac{3}{8} \left(3 q^2+3 q+4\right),\\
&c_1(q)=-\frac{1}{2} q (q+1),
\end{align}
and
\begin{align}
&w_{\frac{3}{2}}(q)=\sqrt{3} \sqrt{q^2+q+1},\\
&w_{\frac{1}{2}}(q)=q^2+q+2.
\end{align}

\subsubsection{$s=2$}
We find
\begin{align}
&c_0(q)=\frac{2 \left(q^5+2 q^4+3 q^3+5 q^2+4 q+3\right)}{3 (q+1)},\\
&c_1(q)=\frac{q \left(-5 q^4-2 q^3+q^2-9q+3\right)}{6 (q+1)},\\
&c_2(q)=\frac{q \left(q^4-q^2+q-1\right)}{6 (q+1)},
\end{align}
and
\begin{align}
&w_{2}(q)=\sqrt{4}\sqrt{q^3+q^2+q+1},\\
&w_{1}(q)=\sqrt{\frac{2}{3}} \sqrt{\frac{q^2+q+1}{q+1}} \left(q^3+q^2+q+3\right).
\end{align}

\subsubsection{$s=5/2$}
We find
\begin{align}
&c_0(q)=\frac{5}{128} \left(15 q^6+15 q^5+25q^4+25 q^3+70 q^2+10 q+64\right),\\
&c_1(q)=\frac{1}{48} q \left(-17 q^5-17 q^4+5 q^3+5 q^2-46 q+22\right),\\
&c_2(q)=\frac{1}{24} q \left(q^5+q^4-q^3-q^2+2 q-2\right),
\end{align}
and
\begin{align}
&w_{\frac{5}{2}}(q)=\sqrt{5} \sqrt{q^4+q^3+q^2+q+1},\\
&w_{\frac{3}{2}}(q)=\frac{1}{\sqrt{2}}\sqrt{\frac{q^3+q^2+q+1}{q+1}} \left(q^4+q^3+q^2+q+4\right),\\
&w_{\frac{1}{2}}(q)=\frac{1}{2} \left(q^6+q^5+2 q^4+2 q^3+5 q^2+q+6\right).
\end{align}

\subsubsection{$s=3$}
We find
\begin{align}
&c_0(q)=\frac{3 \left(q^{11}+2 q^{10}+4 q^9+6 q^8+8 q^7+13 q^6+13 q^5+16 q^4+19 q^3+17 q^2+11 q+10\right)}{10 \left(q^2+q+1\right)},\\
&c_1(q)= -\frac{q \left(49 q^{11}+75 q^{10}+78 q^9+58 q^8+56 q^7+219 q^6+68 q^5-55 q^4+95q^3+212 q^2-130 q-5\right)}{120 (q+1) \left(q^2+q+1\right)},\\
&c_2(q)= \frac{q \left(7 q^{11}+8 q^{10}+3 q^9-8 q^8-9 q^7+21 q^6-2 q^5-23 q^4-7q^3+24 q^2-18 q+4\right)}{60 (q+1) \left(q^2+q+1\right)},\\
&c_3(q)= -\frac{q \left(q^{11}+q^{10}-2 q^8-2 q^7+3 q^6-3 q^4-q^3+4 q^2-2 q+1\right)}{120(q+1) \left(q^2+q+1\right)},
\end{align}
and
\begin{align}
&w_3(q)= \sqrt{6} \sqrt{q^5+q^4+q^3+q^2+q+1},\\
&w_2(q)= \sqrt{\frac{2}{5}} \sqrt{\frac{q^4+q^3+q^2+q+1}{q+1}}\left(q^5+q^4+q^3+q^2+q+5\right),\\
&w_1(q)= \frac{\sqrt{3} \left(q^{11}+2 q^{10}+4 q^9+6 q^8+8 q^7+13 q^6+13 q^5+16 q^4+19 q^3+17 q^2+11q+10\right)}{5 \sqrt{q^5+2 q^4+3 q^3+3 q^2+2 q+1}}.
\end{align}

We remark that in all the above examples the local Hamiltonian $\hat{H}_i(q)$ is positive semi-definite for $q \ge 0$. The proof is as follows. Since $\hat{H}_i(q)$ commutes with ${\hat S}^z_{i,i+1} \coloneqq {\hat s}^z_i + {\hat s}^z_{i+1}$, it is block-diagonal with respect to the eigenspaces of ${\hat S}^z_{i,i+1}$. Let ${\cal V}_M$ be the eigenspace spanned by the basis states $|m_i\rangle_i |m_{i+1}\rangle_{i+1}$ with $m_i+m_{i+1}=M$. Since $w_m (q) > 0$ for all $m$, it is clear that the restriction of $\hat{H}_i(q)$ to ${\cal V}_M$ is $(i)$ a matrix whose off-diagonal entries are all nonpositive and $(ii)$ irreducible, in the basis we work with. Thus we can apply the Perron-Frobenius theorem to conclude that the ground state within ${\cal V}_M$ is unique and can be written as
\begin{align}
|\Psi_M \rangle_{i,i+1} = \sum_{ ( m_i, m_{i+1} ) | m_i+m_{i+1}=M} c_{m_i, m_{i+1}} |m_i\rangle_i |m_{i+1}\rangle_{i+1},
\end{align}
where $c_{m_i, m_{i+1}}>0$ for all $(m_i, m_{i+1})$ such that $m_i+m_{i+1}=M$.

By the construction, the zero-energy state of $\hat{H}_i(q)$ in ${\cal V}_M$ is given as the projection of the local coherent state onto a sector with magnetization $M$ denoted as ${\cal P}_M |\alpha(q) \rangle_{i,i+1}$, where
\begin{align}
    |\alpha (q)\rangle_{i,i+1} & = {\hat {\cal M}
}_i (q) {\hat {\cal M}}_{i+1} (q)\, |\alpha\rangle^{({\rm FM})}_{i,i+1} 
     =  {\hat {\cal M}}_i (q) {\hat {\cal M}}_{i+1} (q)\, e^{\alpha ({\hat s}^-_i + {\hat s}^-_{i+1})} \, |s\rangle_i |s\rangle_{i+1}.
\end{align}
Since coefficients of ${\cal P} |\alpha(q) \rangle_{i,i+1}$ in the $z$-basis are all positive as well, $\langle \Psi_M | {\cal P}_M|\alpha(q) \rangle_{i,i+1} > 0$.
Since the Perron-Frobenius theorem tells that there can be no other eigenstate having only positive coefficients,
$|\Psi_M \rangle_{i,i+1}$ must coincide with the zero-energy state ${\cal P}_M |\alpha(q) \rangle_{i,i+1}$ apart from an overall factor. Since the groundstate energy is zero, the eigenvalues of $\hat{H}_i(q)$ are nonnegative in each ${\cal V}_M$, which proves that $\hat{H}_i(q)$ is positive semi-definite.

\subsection{Witten's conjugation I}
The interpolating Hamiltonian constructed above is inspired by Witten's conjugation~\cite{WITTEN1982, Wouters2021}. Here we explain the connection.

The local Hamiltonian 
\begin{align}
\hat{H}_i^{\text{(FM)}}=Js^2-J\hat{\vec{s}}_i\cdot\hat{\vec{s}}_{i+1}=Js(2s+1)-\frac{J}{2}(\hat{\vec{s}}_i+\hat{\vec{s}}_{i+1})^2
\end{align}
can be decomposed into the summation of the projector onto the spin-$s'$ subspace ($s'=0,1,\cdots,2s$):
\begin{align}
\hat{H}_i^{\text{(FM)}}=J\sum_{s'=0}^{2s}\frac{(2s-s')(2s+1+s')}{2}\hat{P}_{s'}.
\end{align}
Hence, $\hat{H}_i^{\text{(FM)}}$ can be written as 
\begin{align}
&\hat{H}_i^{\text{(FM)}}=J\hat{L}_i^2=J\hat{L}_i^\dagger\hat{L}_i,\\
&\hat{L}_i\coloneqq\sum_{s'=0}^{2s}\sqrt{\frac{(2s-s')(2s+1+s')}{2}}\hat{P}_{s'}.
\end{align}

Now we construct a one-parameter family of Hamiltonians of the following form:
\begin{align}
\hat{H}'(q)&\coloneqq\sum_{i=1}^L\hat{H}_i'(q),\\
\hat{H}_i'(q)&=\hat{\mathcal{M}}(q)^{-1}\hat{L}_i^\dagger\hat{\mathcal{M}}(q)\hat{\mathcal{C}}_i(q)\hat{\mathcal{M}}(q)\hat{L}_i\hat{\mathcal{M}}(q)^{-1}.
\end{align}
for which $|\alpha(q)\rangle$ in Eq.\eqref{alphaq} is a groundstate regardless of $\alpha\in\mathbb{C}$. Here, $\hat{\mathcal{C}}_i(q)$ describes a local operator around the $i$th spin, $\hat{\mathcal{C}}_i(q)$ which is assumed to be positive definite. As far as $\hat{\mathcal{C}}_i(q=0)=1$ and $\hat{H}_i'(q=1)=\hat{H}_i^{\text{(chain)}}$, $\hat{H}_i'(q)$ can be used as an interpolating Hamiltonian. 

For example, for $s=1$, 
\begin{align}
\hat{\mathcal{C}}_i(q)=\frac{3(1+q)^2}{3+4q}-\frac{q(1+q)}{3+4q}\frac{1}{2}(\hat{s}_i^+\hat{s}_{i+1}^-+\text{h.c.})+\frac{q(2+3q)}{3+4q}\hat{s}_i^z\hat{s}_{i+1}^z.
\end{align}
satisfies all the assumptions including the positive definite property for $0\leq q\leq 1$. However, we find that $\hat{\mathcal{C}}_i(q)$ can become quite complicated for higher $s$. To avoid such complication, we directly constructed the interpolating Hamiltonian without constructing $\hat{\mathcal{C}}_i(q)$.

\subsection{Witten's conjugation II}
In this subsection, we discuss another class of frustration-free Hamiltonians obtained by Witten's conjugation. We start with the $s=1/2$ ferromagnetic Heisenberg model $\hat{H}^{\text{(FM)}} = \sum_{i=1}^L\hat{H}_i^{\text{(FM)}}$ with the local Hamiltonian
\begin{align}
& \hat{H}_i^{\text{(FM)}} = J \hat{L}^2_i = J \hat{L}^\dagger_i \hat{L}_i, \\
& {\hat L}_i \coloneqq \frac{1}{4} - \hat{{\bm s}}_i \cdot \hat{{\bm s}}_{i+1},
\end{align}
where the number of sites $L$ is even and $\hat{L}_i$ is a projection operator ($\hat{L}^2_i = \hat{L}_i$). To deform this model, we introduce the operator $\hat{\mathcal{M}}(q)$ via
\begin{align}
    \hat{\mathcal{M}}(q) = q^{\hat{s}^z_1/2} q^{-\hat{s}^z_2/2} \cdots q^{-(-1)^i \hat{s}^z_i/2} \cdots q^{-\hat{s}^z_L/2},
\end{align}
with the inverse
\begin{align}
    \hat{\mathcal{M}}(q)^{-1} = q^{-\hat{s}^z_1/2} q^{\hat{s}^z_2/2} \cdots q^{(-1)^i \hat{s}^z_i/2} \cdots q^{\hat{s}^z_L/2}. 
\end{align}
With this $\hat{\mathcal{M}}(q)$, we get
\begin{align}
    \hat{L}''_i \coloneqq \hat{\mathcal{M}}(q) \hat{L}_i \hat{\mathcal{M}}(q)^{-1}
    = \frac{1}{4} -\frac{1}{2} q^{-(-1)^i} \hat{s}^+_i \hat{s}^-_{i+1} -\frac{1}{2} q^{(-1)^i} \hat{s}^-_i \hat{s}^+_{i+1} -\hat{s}^z_i \hat{s}^z_{i+1},
\end{align}
where we used $\hat{\mathcal{M}}(q) \hat{s}^\pm_i \hat{\mathcal{M}}(q)^{-1}=q^{\mp (-1)^i/2}\hat{s}^\pm_i$ and $\hat{\mathcal{M}}(q) \hat{s}^z_i \hat{\mathcal{M}}(q)^{-1}=\hat{s}^z_i$. The construction of the deformed model proceeds along the same lines as in the previous subsection. We define a one-parameter family of Hamiltonians by
\begin{align}
    & \hat{H}^{\text{(XXZ)}} \coloneqq \sum^L_{i=1} \hat{H}^{\text{(XXZ)}}_i, \label{eq:XXZ_total_Ham}\\
    & \hat{H}^{\text{(XXZ)}}_i \coloneqq (\hat{L}''_i)^\dagger \hat{L}''_i.
\end{align}
A straightforward calculation shows that 
\begin{align}
    \hat{H}^{\text{(XXZ)}}_i = \frac{q+q^{-1}}{2} \left[
    -\hat{s}^x_i \hat{s}^x_{i+1} - \hat{s}^y_i \hat{s}^y_{i+1} + \frac{q+q^{-1}}{2} \left( \frac{1}{4} - \hat{s}^z_i \hat{s}^z_{i+1}\right) + \frac{q-q^{-1}}{4} (-1)^i (\hat{s}^z_i - \hat{s}^z_{i+1})
    \right],
    \label{eq:XXZ_stag}
\end{align}
which is, up to a prefactor, the local Hamiltonian of the frustration-free model studied in previous work \cite{alcaraz1995critical, karadamoglou1999bulk, wouters2018exact}.

Since Witten's conjugation does not change the number of zero-energy states, the ground state degeneracy of $\hat{H}^{\text{(XXZ)}}$ is $L+1$-fold. By construction it is clear that these degenerate ground states are obtained by acting with a deformed lowering operator on the fully polarized state $|\Phi_0\rangle = \bigotimes^L_{i=1}|\frac{1}{2}\rangle_i$:
\begin{align}
    |M\rangle^{(\text{XXZ})} \coloneqq \frac{1}{N''_M}  (\hat{\mathcal{S}}^-)^{\frac{L}{2}-M} |\Phi_0 \rangle, \quad M=-\frac{L}{2}, -\frac{L}{2}+1, \cdots, \frac{L}{2},
\end{align}
where $\hat{\mathcal{S}}^- = \hat{\mathcal{M}}(q) \hat{S}^- \hat{\mathcal{M}}(q)^{-1}$ is the deformed lowering operator and $N''_M$ is a normalization prefactor.

Interestingly, the $s=1/2$ model in Eq. (\ref{eq:XXZ_stag}) is deeply connected to a seemingly unrelated model -- an anisotropic extension of the $s=1$ ferromagnetic biquadratic model \cite{shi2022instability, zhou2023alternative}. The Hamiltonian of the model is given by
\begin{align}
    & \hat{H}^{(\text{BQ})} (J_x, J_y, J_z) \coloneqq \sum^L_{i=1} \hat{H}^{(\text{BQ})}_i (J_x, J_y, J_z), \\
    & \hat{H}^{(\text{BQ})}_i (J_x, J_y, J_z) \coloneqq (J_x \hat{s}^x_i \hat{s}^x_{i+1} + J_y \hat{s}^y_i \hat{s}^y_{i+1} + J_z \hat{s}^z_i \hat{s}^z_{i+1} )^2, 
\end{align}
where the number of sites $L$ is even. In the following, we focus on the special line in the parameter space where $J_x/J_y > 0$ and $J_z=0$, and argue that the model in a particular subspace can be mapped to the $s=1/2$ model (\ref{eq:XXZ_total_Ham}). Note that the same argument applies to the lines obtained by cyclic permutations of the indices $x$, $y$, and $z$.

To proceed, we first note that the subspace consisting of $|\psi\rangle$ such that $\hat{s}^z_i |\psi\rangle = \pm |\psi\rangle$ for all $i$ is an invariant subspace of the Hamiltonian $\hat{H}^{(\text{BQ})} (J_x, J_y, 0)$. In other words, $\hat{H}^{(\text{BQ})}_i (J_x, J_y, 0)$ does not create $|0\rangle$ when acting on a product state of the form $|m_i\rangle_i |m_{i+1} \rangle_{i+1}$, where $m_i m_{i+1}=\pm 1$. 
We denote by $\mathcal{W}$ the invariant subspace and introduce effective Pauli operators:
\begin{align}
    \hat{\sigma}^+_i = |1\rangle_i \langle -1|, \quad
    \hat{\sigma}^-_i = |-1 \rangle_i \langle 1|, \quad
    \hat{\sigma}^z_i = |1 \rangle_i \langle 1| - |-1\rangle_i \langle -1|.
\end{align}
Let us denote by $\hat{\mathcal{P}}$ the orthogonal projection onto $\mathcal{W}$. A tedious but straightforward calculation shows that
\begin{align}
    \hat{H}^{(\text{eff})} (J_x, J_y, 0) & =
    \hat{\mathcal{P}} \hat{H}^{(\text{BQ})} (J_x, J_y, 0)
    \hat{\mathcal{P}} =
    \sum^L_{i=1} \hat{H}^{(\text{eff})}_i (J_x, J_y, 0), \\
    \hat{H}^{(\text{eff})}_i (J_x, J_y, 0) & = \sum^L_{i=1} \left[
    \frac{J^2_x+J^y_2}{4} \hat{\sigma}^x_i \hat{\sigma}^x_{i+1} + \frac{J_x J_y}{2} \hat{\sigma}^y_i \hat{\sigma}^y_{i+1} - \frac{J_x J_y}{2} \hat{\sigma}^z_i \hat{\sigma}^z_{i+1} + \frac{J^2_x-J^2_y}{4} (\hat{\sigma}^x_i + \hat{\sigma}^x_{i+1})+\frac{J^2_x + J^2_y}{4}
    \right],
    \label{eq:XXZ_stag1}
\end{align}
where $\hat{\sigma}^x_i = \hat{\sigma}^+_i + \hat{\sigma}^-_i$ and $\hat{\sigma}^y_i = (\hat{\sigma}^+_i - \hat{\sigma}^-_i)/\mathrm{i}$. To make the link to the Hamiltonian in Eq. (\ref{eq:XXZ_stag}), we perform a unitary transformation such that $\hat{\sigma}^x_i \to (-1)^i \hat{\sigma}^z_i$, $\hat{\sigma}^y_i \to (-1)^i \hat{\sigma}^x_i$, and $\hat{\sigma}^z_i \to \hat{\sigma}^y_i$ in Eq. (\ref{eq:XXZ_stag1}). As a result, we obtain the local Hamiltonian of $\hat{H}^{(\text{eff})} (J_x, J_y, 0)$ in the new basis:
\begin{align}
    \hat{\widetilde{H}}^{(\text{eff})}_i (J_x, J_y, 0) = \frac{J_x J_y}{2} \left[
    -\hat{\sigma}^x_i \hat{\sigma}^x_{i+1} - \hat{\sigma}^y_i \hat{\sigma}^y_{i+1} + \frac{\frac{J_x}{J_y}+\frac{J_y}{J_x}}{2} (1-\hat{\sigma}^z_i \hat{\sigma}^z_{i+1}) + (-1)^i \frac{\frac{J_x}{J_y}-\frac{J_y}{J_x}}{2} (\hat{\sigma}^z_i - \hat{\sigma}^z_{i+1})
    \right].
\end{align}
With the identification $J_x/J_y \leftrightarrow q$, this local Hamiltonian is, up to a multiplicative constant, equivalent to $\hat{H}^{\text{(XXZ)}}_i$ in Eq. (\ref{eq:XXZ_stag}). This correspondence between the $s=1$ and $s=1/2$ models clearly explains the presence of highly degenerate ground states on the special line $J_x/J_y > 0$ and $J_z=0$ found in Refs. \cite{shi2022instability, zhou2023alternative}. Moreover, since the ground states of $\hat{H}^{(\text{BQ})} (J_x, J_y, 0)$ are annihilated by $J_x \hat{s}^x_i \hat{s}^x_{i+1} + J_y \hat{s}^y_i \hat{s}^y_{i+1}$ for all $i$, they are zero-energy ground states of a more general Hamiltonian
\begin{align}
    \hat{H}^{(2n)} (J_x, J_y, 0) \coloneqq \sum^L_{i=1} (J_x \hat{s}^x_i \hat{s}^x_{i+1} + J_y \hat{s}^y_i \hat{s}^y_{i+1})^{2n},
\end{align}
where $n=2, 3, \cdots$. This is in accord with a recent study by Y-W. Dai et al. \cite{dai2023ground}, where the $n=2$ case was studied in detail.

\section{The Bogoliubov inequality} \label{sec:HMW}
\subsection{$T>0$}

To derive the Bogoliubov inequality, we introduce the correlation function of two operators $\hat{A}$ and $\hat{B}$:
\begin{align}
(\hat{A},\hat{B})= (\hat{B}^\dagger,\hat{A}^\dagger)  := \beta\int_0^1dx\Big\langle e^{\beta x\hat{H}}\hat{A}e^{-\beta x\hat{H}}\hat{B}^\dagger \Big\rangle=\sum_{n,m}\langle n|\hat{A}|m\rangle \langle m|\hat{B}^\dagger |n\rangle \frac{e^{-\beta E_m}-e^{-\beta E_n}}{Z}\frac{1}{E_n-E_m},
\end{align}
where the expectation value is taken by the Gibbs state of $\hat{H}$. Since this operation is a well-defined inner product between two operators $\hat{A}$ and $\hat{B}$, the Cauchy--Schwarz inequality tells that $(\hat{A},\hat{A})(\hat{B},\hat{B}) \geq |(\hat{A},\hat{B})|^2$. 
By replacing $\hat{H}$ with $\hat{H}(h)$ and plugging in $\hat{A} = \hat{X}_{\bm{k}}$ and $\hat{B} = [\hat{Q}_{\bm{k}}, \hat{H}(h)]$, we obtain 
\begin{align}
&(\hat{A},\hat{A}) := \beta\int_0^1dx\Big\langle e^{\beta x\hat{H}(h)}\hat{X}_{\bm{k}}e^{-\beta x\hat{H}(h)}\hat{X}_{\bm{k}}^\dagger \Big\rangle,\\
&(\hat{A},\hat{B}) := \beta\int_0^1dx\Big\langle e^{\beta x\hat{H}(h)}\hat{X}_{\bm{k}}e^{-\beta x\hat{H}(h)}[\hat{H}(h),\hat{Q}_{\bm{k}}^\dagger] \Big\rangle=\langle[\hat{X}_{\bm{k}},\hat{Q}_{\bm{k}}^\dagger]\rangle,\\
&(\hat{B},\hat{B}) := \beta\int_0^1dx\Big\langle e^{\beta x\hat{H}(h)}[\hat{Q}_{\bm{k}},\hat{H}(h)]e^{-\beta x\hat{H}(h)}[\hat{H}(h),\hat{Q}_{\bm{k}}^\dagger] \Big\rangle=\langle[[\hat{Q}_{\bm{k}},\hat{H}(h)],\hat{Q}_{\bm{k}}^\dagger]\rangle.
\end{align}
Also, we have
\begin{align}
&(\hat{A},\hat{A})=\sum_{n,m}|\langle n|\hat{A}|m\rangle|^2\frac{e^{-\beta E_m}-e^{-\beta E_n}}{Z}\frac{1}{E_n-E_m}\leq \frac{\beta}{2}\sum_{n,m}|\langle n|\hat{A}|m\rangle|^2 \frac{e^{-\beta E_m}+e^{-\beta E_n}}{Z}=\frac{\beta}{2}\langle\hat{A}\hat{A}^\dagger+\hat{A}^\dagger\hat{A}\rangle,
\end{align}
where we used
\begin{align}
\frac{e^{-\beta E_m}-e^{-\beta E_n}}{E_n-E_m}<\frac{\beta}{2}(e^{-\beta E_m}+e^{-\beta E_n}),
\end{align}
which follows from $\tanh(x)\leq x$. Therefore, the Cauchy--Schwarz inequality gives
\begin{align}
\langle\hat{X}_{\bm{k}}\hat{X}_{\bm{k}}^\dagger+\hat{X}_{\bm{k}}^\dagger\hat{X}_{\bm{k}}\rangle \geq \frac{2T|\langle[i\hat{Q}_{\bm{k}}^\dagger,\hat{X}_{\bm{k}}]\rangle|^2}{\langle[[\hat{Q}_{\bm{k}},\hat{H}(h)],\hat{Q}_{\bm{k}}^\dagger]\rangle}.
\end{align}

\subsection{$T=0$}

To obtain the zero-temperature version of the Bogoliubov inequality, we define the following operator inner product:
\begin{align}
    (\hat{A},\hat{B}) := \Big\langle \hat{A} \frac{\hat{P}}{\hat{H} - E_\textrm{GS}} \hat{B}^\dagger + \hat{B}^\dagger \frac{\hat{P}}{\hat{H} - E_\textrm{GS}} \hat{A} \Big\rangle,
\end{align}
where the expectation value is taken in the groundstate $| \Phi_\textrm{GS} \rangle$ of $\hat{H}$, $E_\textrm{GS}$ is the groundstate energy, $\hat{P}$ is the projection onto excited states. 
By plugging in $\hat{A} = \hat{X}_{\bm{k}}$ and $\hat{B} = [\hat{Q}_{\bm{k}}, \hat{H}(h)]$, we obtain 
\begin{align}
\Big\langle\hat{X}_{\bm{k}}\frac{\hat{P}}{\hat{H}(h)-E_{\mathrm{GS}}}\hat{X}_{\bm{k}}^\dagger+\hat{X}_{\bm{k}}^\dagger\frac{\hat{P}}{\hat{H}(h)-E_{\mathrm{GS}}}\hat{X}_{\bm{k}}\Big\rangle  \langle[\hat{Q}_{\bm{k}},[\hat{H}(h),\hat{Q}_{\bm{k}}^\dagger]]\rangle
\geq \big| \big\langle\hat{X}_{\bm{k}}\hat{P} \hat{Q}_{\bm{k}}^\dagger-  \hat{Q}_{\bm{k}}^\dagger\hat{P}\hat{X}_{\bm{k}}\big\rangle \big|^2 = \big|\langle[\hat{X}_{\bm{k}},\hat{Q}_{\bm{k}}^\dagger]\rangle\big|^2. \label{BI}
\end{align}
By the definition of $\omega_{\bm{k}}(h)$, we find
\begin{align}
\Big\langle\hat{X}_{\bm{k}}\frac{\hat{P}}{\hat{H}(h)-E_{\mathrm{GS}}}\hat{X}_{\bm{k}}^\dagger+\hat{X}_{\bm{k}}^\dagger\frac{\hat{P}}{\hat{H}(h)-E_{\mathrm{GS}}}\hat{X}_{\bm{k}}\Big\rangle
\leq\frac{1}{\omega_{\bm{k}}(h)}\Big\langle\hat{X}_{\bm{k}}\hat{P}\hat{X}_{\bm{k}}^\dagger+\hat{X}_{\bm{k}}^\dagger\hat{P}\hat{X}_{\bm{k}}\Big\rangle\leq\frac{1}{\omega_{\bm{k}}(h)}\Big\langle\hat{X}_{\bm{k}}\hat{X}_{\bm{k}}^\dagger+\hat{X}_{\bm{k}}^\dagger\hat{X}_{\bm{k}}\Big\rangle.\label{omegaineq}
\end{align}
Combining this with the Bogoliubov inequality \eqref{BI}, we obtain the expression we used in the main text: 
\begin{align}
\langle\hat{X}_{\bm{k}}\hat{X}_{\bm{k}}^\dagger+\hat{X}_{\bm{k}}^\dagger\hat{X}_{\bm{k}}\rangle \geq\frac{\omega_{\bm{k}}(h)\big|\langle[i\hat{Q}_{\bm{k}}^\dagger,\hat{X}_{\bm{k}}]\rangle\big|^2}{ \langle[\hat{Q}_{\bm{k}},[\hat{H}(h),\hat{Q}_{\bm{k}}^\dagger]]\rangle}. 
\end{align}

\section{The Nambu--Goldstone theorem} \label{sec:NG}

The Nambu--Goldstone theorem guarantees the presence of a gapless excitation when a continuous symmetry is spontaneously broken, i.e., $m(0^+)=\lim_{h\to0^+}\lim_{V\to\infty}m(h)\neq0$. Here we review the proof based on the Bogoliubov inequality, following Ref.~\onlinecite{Wagner,Stringari}. 
To this end, instead of inequality in \eqref{omegaineq}, we use
\begin{align}
\Big\langle\hat{X}_{\bm{k}}\frac{\hat{P}}{\hat{H}(h)-E_{\mathrm{GS}}}\hat{X}_{\bm{k}}^\dagger+\hat{X}_{\bm{k}}^\dagger\frac{\hat{P}}{\hat{H}(h)-E_{\mathrm{GS}}}\hat{X}_{\bm{k}}\Big\rangle
&\leq\frac{1}{\omega_{\bm{k}}(h)^2}\langle\hat{X}_{\bm{k}}(\hat{H}(h)-E_{\mathrm{GS}})\hat{X}_{\bm{k}}^\dagger+\hat{X}_{\bm{k}}^\dagger(\hat{H}(h)-E_{\mathrm{GS}})\hat{X}_{\bm{k}}\rangle\notag\\
&=\frac{1}{\omega_{\bm{k}}(h)^2}\langle[\hat{X}_{\bm{k}},[\hat{H}(h),\hat{X}_{\bm{k}}^\dagger]]\rangle.
\end{align}
Combining this with  the Bogoliubov inequality \eqref{BI}, we find
\begin{align}
\omega_{\bm{k}}(h)^2\leq\frac{\langle[\hat{X}_{\bm{k}},[\hat{H}(h),\hat{X}_{\bm{k}}^\dagger]]\rangle  \langle[\hat{Q}_{\bm{k}}^\dagger,[\hat{H}(h),\hat{Q}_{\bm{k}}]]\rangle\notag}{ \big|\langle[i\hat{Q}_{\bm{k}}^\dagger,\hat{X}_{\bm{k}}]\rangle\big|^2}.
\end{align}
For $|\bm{k} | \ll 1$, one can take 
\begin{align}
&\lim_{h\to0^+}\lim_{V\to\infty}\frac{1}{V}\langle[i\hat{Q}^\dagger_{ \bm{k} }, \hat{X}_{\bm{k}} ]\rangle= m(0^+), \\
&\lim_{h\to0^+}\lim_{V\to\infty}\frac{1}{V} \langle[\hat{Q}_{\bm{k}}^\dagger,[\hat{H}(h),\hat{Q}_{\bm{k}}]]\rangle \sim C | \bm{k} |^{2 n_0}, \\
&\lim_{h\to0^+}\lim_{V\to\infty}\frac{1}{V} \langle[\hat{X}_{\bm{k}},[\hat{H}(h),\hat{X}_{\bm{k}}^\dagger]]\rangle \sim C'. 
\end{align}
Then, we can show that 
\begin{align}
\lim_{h\to0^+}\lim_{V\to\infty}\omega_{\bm{k}}(h)&\leq \sqrt{\frac{ C' C}{|m(0^+)|^2}}|\bm{k}|^{n_0}.
\end{align}

\section{Theorems on frustration-free Hamiltonians} \label{sec:FF}
Here, following Refs.~\onlinecite{Knabe,GossetMozgunov}, we review the derivation of the bound for excitation energies in frustration-free translation-invariant Hamiltonians.

Suppose that $\hat{H}=\sum_{i=1}^L\hat{H}_i$ is frustration-free. We assume the invariance under translation operator $\hat{T}$, i.e., $\hat{T}\hat{H}_i\hat{T}^{\dagger}=\hat{H}_{i+1}$ for all $i$. Without loss of the generality, we can assume that (i) $\hat{H}_i$ acts nontrivially only on the site $i$ and $i+1$ and (ii) $\hat{H}_i$ is a projection operator ($\hat{H}_i^2=\hat{H}_i$) onto excited states.
If the first assumption is not satisfied, we can combine several sites together (i.e., coarse-graining) until this assumption holds.
Furthermore, if the second assumption is not automatically fulfilled, we modify eigenvalues of $\hat{H}_i$, without changing eigenvectors, in such a way that the smallest eigenvalue is $0$ and other eigenvalues are $1$. This process keeps all essential features such as whether the system is gapless or gapped, as well as the scaling of the gapless excitation. Without loss of generality, after shifting $H_i$ by a proper constant, $\hat{H}_i = \sum_n a_n \mathcal{P}_{i,n}$, where $a_n > 0$ and $\mathcal{P}_{i,n}$ is a projector onto an $n$-th eigenvector. The groundstate of $\hat{H} = \sum_i \hat{H}_i$ has a zero energy. Among non-zero eigenvalues of $\hat{H}_i$, denote $a_\textrm{min/max}$ as the smallest/largest eigenvalue of $\hat{H}_i$. Now, define $\hat{Q}_i := \sum_n \mathcal{P}_{i,n}$ and $\hat{H}^Q:= \sum_i \hat{Q}_i$. Then, due to the positive semidefiniteness of projector operators, it is straightforward that 
\begin{align}
    a_\textrm{min} \gamma_Q(L) \leq \gamma(L) \leq     a_\textrm{max} \gamma_Q(L)
\end{align}
where $\gamma(L)$ and $\gamma_Q(\L)$ are gaps for $\hat{H}$ and $\hat{H}^Q$ respectively for system size $L$. Therefore, by studying $\hat{H}^Q$, one can rigorously study the scaling form of the excitation gap of $\hat{H}$. 

\subsection{Knabe's theorem}
We set the groundstate energy to be $0$. Let us take a length $\ell$ ($3\leq\ell< L/2$) and define subsystem Hamiltonian by 
\begin{align}
\hat{H}_{\ell,x_0}^{\text{OBC}}\,{\coloneqq}\,\sum_{i=0}^{\ell-2}\hat{H}_{x_0+i}.
\end{align}
We write the smallest nonzero eigenvalue of $\hat{H}$ and $\hat{H}_{\ell,x_0}^{\text{OBC}}$ as $\epsilon_L^{\text{PBC}}$ and $\epsilon_\ell^{\text{OBC}}$, respectively. 

Knabe showed~\cite{Knabe}
\begin{align}
&\epsilon_L^{\text{PBC}}\geq\frac{\ell-1}{\ell-2}\left(\epsilon_\ell^{\text{OBC}}-\frac{1}{\ell-1}\right).\label{Knabe}
\end{align}
This relation can be shown by the property $\hat{H}_i^2=\hat{H}_i$ of the projector. 
\begin{align}
&\hat{H}^2+\frac{1}{\ell-2}\hat{H}-\frac{1}{\ell-2}\sum_{x_0=1}^L(\hat{H}_{\ell,x_0}^{\text{OBC}})^2\notag\\
&=\sum_{i,j=1}^L\hat{H}_i\hat{H}_j+\frac{1}{\ell-2}\hat{H}-\frac{1}{\ell-2}\sum_{x_0=1}^L\sum_{i,j=0}^{\ell-2}\hat{H}_{x_0+i}\hat{H}_{x_0+j}\notag\\
&=\left(\hat{H}+2\sum_{i=1}^L\hat{H}_i\hat{H}_{i+1}+\sum_{1\leq i,j\leq L,\,2\leq |i-j|}\hat{H}_i\hat{H}_j\right)+\frac{1}{\ell-2}\hat{H}\notag\\
&\quad-\frac{1}{\ell-2}\left(\sum_{x_0=1}^L\sum_{i=0}^{\ell-2}\hat{H}_{x_0+i}+2\sum_{x_0=1}^L\sum_{i=0}^{\ell-3}\hat{H}_{x_0+i}\hat{H}_{x_0+i+1}+\sum_{x_0=1}^L\sum_{0\leq i,j\leq\ell-2,\,2\leq |i-j|}\hat{H}_{x_0+i}\hat{H}_{x_0+j}\right)\notag\\
&=\sum_{1\leq i,j\leq L, \,2\leq|i-j|}\hat{H}_i\hat{H}_j-\frac{1}{\ell-2}\sum_{x_0=1}^L\sum_{0\leq i,j\leq\ell-2,\,2\leq |i-j|}\hat{H}_{x_0}\hat{H}_{x_0+(j-i)}\notag\\
&=\sum_{0\leq i,j\leq L,\,\ell-1\leq |i-j|}\hat{H}_i\hat{H}_j+2\sum_{i=1}^L\sum_{d=2}^{\ell-2}\hat{H}_i\hat{H}_{i+d}-\frac{2}{\ell-2}\sum_{x_0=1}^L\sum_{d=2}^{\ell-2}\left(\sum_{i,j=0}^{\ell-2}\delta_{d,j-i}\right)\hat{H}_{x_0}\hat{H}_{x_0+d}\notag\\
&=\sum_{0\leq i,j\leq L,\,\ell-1\leq |i-j|}\hat{H}_i\hat{H}_j+2\sum_{d=2}^{\ell-2}\frac{d-1}{\ell-2}\sum_{i=1}^L\hat{H}_i\hat{H}_{i+d}\notag\\
&\geq0.
\end{align}
In the second last line, we used $\sum_{i,j=0}^{\ell-2}\delta_{d,j-i}=\ell-1-d$.
The last inequality follows because the products of commuting projectors are positive semidefinite.  

Also, we have
\begin{align}
\sum_{x_0=1}^L(\hat{H}_{\ell,x_0}^{\text{OBC}})^2\geq \epsilon_\ell^{\text{OBC}}\sum_{x_0=1}^L\hat{H}_{\ell,x_0}^{\text{OBC}}=(\ell-1)\epsilon_\ell^{\text{OBC}}\hat{H}.
\end{align}
Therefore,
\begin{align}
&\hat{H}^2+\frac{1}{\ell-2}\hat{H}\geq\frac{1}{\ell-2}\sum_{x_0=1}^L(\hat{H}_{\ell,x_0}^{\text{OBC}})^2\geq\frac{\ell-1}{\ell-2}\epsilon_\ell^{\text{OBC}}\hat{H}\notag\\
&\hat{H}^2\geq\frac{\ell-1}{\ell-2}\left(\epsilon_\ell^{\text{OBC}}-\frac{1}{\ell-1}\right)\hat{H}.
\end{align}
This gives the inequality \eqref{Knabe}.

\subsection{Gosset--Mozgunov's improvement}
Gosset--Mozgunov improved the bound in \eqref{Knabe} by modifying the OBC Hamiltonian:
\begin{align}
\hat{\tilde{H}}_{\ell,x_0}^{\text{OBC}}\coloneqq\sum_{i=0}^{\ell-2}c_i\hat{H}_{x_0+i}
\end{align}
with parameters $c_i=(\ell-1)+(\ell-2)i-i^2=\frac{\ell^2}{4}-(i+1-\frac{\ell}{2})^2$. 
We will use
\begin{align}
\sum_{i=0}^{\ell-3}c_ic_{i+1}=\frac{\ell(\ell^2-1)(\ell^2-4)}{30},\quad \sum_{i=0}^{\ell-2}c_i^2=\frac{\ell(\ell^4-1)}{30},\quad \sum_{i=0}^{\ell-2}c_i=\frac{\ell(\ell^2-1)}{6}.
\end{align}
This time we find
\begin{align}
&\hat{H}^2+\beta\hat{H}-\alpha\sum_{x_0=1}^L(\hat{\tilde{H}}_{\ell,x_0}^{\text{OBC}})^2\notag\\
&=\sum_{i,j=1}^L\hat{H}_i\hat{H}_j+\beta\hat{H}-\alpha\sum_{x_0=1}^L\sum_{i,j=0}^{\ell-2}c_ic_j\hat{H}_{x_0+i}\hat{H}_{x_0+j}\notag\\
&=\left(\hat{H}+2\sum_{i=1}^L\hat{H}_i\hat{H}_{i+1}+\sum_{0\leq i,j\leq L,\,2\leq |i-j|}\hat{H}_i\hat{H}_j\right)+\beta\hat{H}\notag\\
&\quad-\alpha\left(\sum_{x_0=1}^L\sum_{i=0}^{\ell-2}c_i^2\hat{H}_{x_0+i}+2\sum_{x_0=1}^L\sum_{i=0}^{\ell-3}c_ic_{i+1}\hat{H}_{x_0+i}\hat{H}_{x_0+i+1}+\sum_{x_0=1}^L\sum_{0\leq i,j\leq\ell-2,\,2\leq |i-j|}c_ic_j\hat{H}_{x_0+i}\hat{H}_{x_0+j}\right)\notag\\
&=\left(1+\beta-\alpha\sum_{i=0}^{\ell-2}c_i^2\right)\hat{H}+2\left(1-\alpha\sum_{i=0}^{\ell-3}c_ic_{i+1}\right)\sum_{j=1}^L\hat{H}_j\hat{H}_{j+1}\notag\\
&\quad+\sum_{0\leq i,j\leq L,\,2\leq |i-j|}\hat{H}_i\hat{H}_j-\alpha\sum_{x_0=1}^L\sum_{0\leq i,j\leq\ell-2,\,2\leq |i-j|}c_ic_j\hat{H}_{x_0+i}\hat{H}_{x_0+j}\notag\\
&=\sum_{0\leq i,j\leq L,\,\ell-1\leq |i-j|}\hat{H}_i\hat{H}_j+2\sum_{d=2}^{\ell-2}\frac{\sum_{i=0}^{\ell-3}c_ic_{i+1}-\sum_{i=0}^{\ell-2-d}c_ic_{i+d}}{\sum_{i=0}^{\ell-3}c_ic_{i+1}}\sum_{j=1}^L\hat{H}_j\hat{H}_{j+d}\notag\\
&=\sum_{0\leq i,j\leq L,\,\ell-1\leq |i-j|}\hat{H}_i\hat{H}_j+2\sum_{d=2}^{\ell-2}\frac{(d^2-1)(5\ell^3-5d\ell^2-5\ell+d^3+d)}{\ell(\ell^2-1)(\ell^2-4)}\sum_{j=1}^L\hat{H}_j\hat{H}_{j+d}\notag\\
&\geq0
\end{align}
In the derivation, we set $\alpha$ and $\beta$ in such a way that
\begin{align}
1-\alpha\sum_{i=0}^{\ell-3}c_ic_{i+1}=0,\quad1+\beta-\alpha\sum_{i=0}^{\ell-2}c_i^2=0,\quad\text{i.e.},\quad\alpha=\frac{30}{\ell(\ell^2-1)(\ell^2-4)},\quad\beta=\frac{5}{\ell^2-4}.
\end{align}
Also, we used the fact that $5\ell^3-5d\ell^2-5\ell+d^3+d>0$ for all $2\leq d\leq \ell-2$.

Let us consider the eigenspace of $\hat{H}$ with eigenvalue $\epsilon_L^{\text{PBC}}$. 
As we show in the next subsection, there exists a state $|\phi\rangle$ in this space such that
\begin{align}
\langle\phi|(\hat{\tilde{H}}_{\ell,x_0}^{\text{OBC}})^2|\phi\rangle\geq\epsilon_\ell^{\text{OBC}}\frac{\ell(\ell+1)}{6}\langle\phi|\hat{\tilde{H}}_{\ell,x_0}^{\text{OBC}}|\phi\rangle.\label{key}
\end{align}
Hence,
\begin{align}
\langle\phi|(\hat{\tilde{H}}_{\ell,x_0}^{\text{OBC}})^2|\phi\rangle\geq\epsilon_\ell^{\text{OBC}}\frac{\ell(\ell+1)}{6}\sum_{i=0}^{\ell-2}c_i\langle\phi|\hat{H}_{x_0+i}|\phi\rangle,
\end{align}
which implies
\begin{align}
\sum_{x_0=1}^L\langle\phi|(\hat{\tilde{H}}_{\ell,x_0}^{\text{OBC}})^2|\phi\rangle\geq \epsilon_\ell^{\text{OBC}}\epsilon_L^{\text{PBC}}\frac{\ell(\ell+1)}{6}\sum_{i=0}^{\ell-2}\hat{c}_i=\frac{\ell^2(\ell+1)^2(\ell-1)}{36}\epsilon_\ell^{\text{OBC}}\epsilon_L^{\text{PBC}}.
\end{align}

Combining these results, we find
\begin{align}
(\epsilon_L^{\text{PBC}})^2+\beta\epsilon_L^{\text{PBC}}\geq\alpha\sum_{x_0=1}^L\langle\phi|(\hat{\tilde{H}}_{\ell,x_0}^{\text{OBC}})^2|\phi\rangle\geq\frac{5\ell(\ell+1)}{6(\ell^2-4)}\epsilon_\ell^{\text{OBC}}\epsilon_L^{\text{PBC}}.
\end{align}
Therefore,
\begin{align}
\epsilon_L^{\text{PBC}}\geq \frac{5\ell(\ell+1)}{6(\ell^2-4)}\epsilon_\ell^{\text{OBC}}-\frac{5}{\ell^2-4}=\frac{5\ell(\ell+1)}{6(\ell^2-4)}\left(\epsilon_\ell^{\text{OBC}}-\frac{6}{\ell(\ell+1)}\right).
\end{align}

\subsection{Proof of Eq.\eqref{key}}
We choose $|\phi\rangle$ to be a simultaneous eigenstate of the Hamiltonian $\hat{H}$ and the translation operator $\hat{T}$ with eigenvalues $\epsilon_L^{\text{PBC}}$ and $e^{i\theta}$, respectively.

Let ${\cal W}_0$ denote the groundstate subspace of $\hat{\tilde{H}}_{\ell,x_0}^{\text{OBC}}$ and let $\hat{P}_{\ell,x_0}^{\text{OBC}}$ denote the projector onto ${\cal W}_0$. Then, we have
\begin{align}
\hat{P}_{\ell,x_0}^{\text{OBC}}\hat{H}_{x_0+i}=\hat{H}_{x_0+i}\hat{P}_{\ell,x_0}^{\text{OBC}}=0
\end{align}
for $0\leq i\leq \ell-2$. It follows that
$\hat{Q}_{\ell,x_0}^{\text{OBC}}\coloneqq 1-\hat{P}_{\ell,x_0}^{\text{OBC}}$, 
the projector onto the orthogonal complement of ${\cal W}_0$, satisfies 
\begin{align}
\hat{Q}_{\ell,x_0}^{\text{OBC}}\hat{H}_{x_0+i}=\hat{H}_{x_0+i}\hat{Q}_{\ell,x_0}^{\text{OBC}}=\hat{H}_{x_0+i}
\end{align}
for $0\leq i\leq \ell-2$. 

If $\hat{Q}_{\ell,x_0}^{\text{OBC}}|\phi\rangle=0$ then Eq.\eqref{key} trivially holds. So we assume $\hat{Q}_{\ell,x_0}^{\text{OBC}}|\phi\rangle\neq0$. Then we define 
\begin{align}
|\phi'\rangle\coloneqq\frac{1}{\|\hat{Q}_{\ell,x_0}^{\text{OBC}}|\phi\rangle\|}\hat{Q}_{\ell,x_0}^{\text{OBC}}|\phi\rangle.
\end{align}

By these definitions, we have
\begin{align}
\langle\phi|(\hat{\tilde{H}}_{\ell,x_0}^{\text{OBC}})^2|\phi\rangle=\langle\phi'|(\hat{\tilde{H}}_{\ell,x_0}^{\text{OBC}})^2|\phi'\rangle\langle\phi|\hat{Q}_{\ell,x_0}^{\text{OBC}}|\phi\rangle\geq\langle\phi'|\hat{\tilde{H}}_{\ell,x_0}^{\text{OBC}}|\phi'\rangle^2\langle\phi|\hat{Q}_{\ell,x_0}^{\text{OBC}}|\phi\rangle=\langle\phi'|\hat{\tilde{H}}_{\ell,x_0}^{\text{OBC}}|\phi'\rangle\langle\phi|\hat{\tilde{H}}_{\ell,x_0}^{\text{OBC}}|\phi\rangle.\label{GMineq1}
\end{align}

Also, by the translation invariance of $|\phi\rangle$, 
\begin{align}
\langle\phi'|\hat{H}_{x_0+i}|\phi'\rangle=\frac{\langle\phi|\hat{Q}_{\ell,x_0}^{\text{OBC}}\hat{H}_{x_0+i}\hat{Q}_{\ell,x_0}^{\text{OBC}}|\phi\rangle}{\langle\phi|\hat{Q}_{\ell,x_0}^{\text{OBC}}|\phi\rangle}=\frac{\langle\phi|\hat{H}_{x_0+i}|\phi\rangle}{\langle\phi|\hat{Q}_{\ell,x_0}^{\text{OBC}}|\phi\rangle}
\end{align}
is independent of $0\leq i\leq \ell-2$. Therefore,
\begin{align}
\langle\phi'|\hat{\tilde{H}}_{\ell,x_0}^{\text{OBC}}|\phi'\rangle
=\sum_{i=0}^{\ell-2}c_i\langle\phi'|\hat{H}_{x_0+i}|\phi'\rangle
=\frac{1}{\ell-1}\sum_{i=0}^{\ell-2}c_i 
\langle\phi'|\hat{\tilde{H}}_{\ell,x_0}^{\text{OBC}}|\phi'\rangle
\geq\frac{\ell(\ell+1)}{6}\epsilon_\ell^{\text{OBC}}.
\label{GMineq2}
\end{align}
Combining \eqref{GMineq1} and \eqref{GMineq2}, we obtain Eq.\eqref{key}.

\end{document}